\documentclass[reprint,superscriptaddress,
showpacs,
 amsmath,amssymb,
 aps,
]{revtex4-1}

\usepackage{graphicx}
\usepackage{dcolumn}
\usepackage{bm}
\usepackage{multirow}
\usepackage{longtable}
\usepackage{rotating}
\usepackage{color}

\begin{document}

\preprint{APS/123-QED}

\title{Design of NbN superconducting nanowire single photon detectors with enhanced infrared photon detection efficiency}

\author{Q. Wang}
\affiliation{Huygens Kamerlingh-Onnes Laboratory, Leiden University, Niels Bohrweg 2, 2333 CA Leiden, The Netherlands}
\affiliation{Physics Institute of the University of Zurich, Winterthurerstr. 190, 8057 Zurich, Switzerland}

\author{J. J. Renema}
\affiliation{Huygens Kamerlingh-Onnes Laboratory, Leiden University, Niels Bohrweg 2, 2333 CA Leiden, The Netherlands}
\affiliation{Clarendon Laboratory, University of Oxford, Parks Road, Oxford OX1 3PU, United Kingdom}

\author{A. Engel}
\affiliation{Physics Institute of the University of Zurich, Winterthurerstr. 190, 8057 Zurich, Switzerland}


\author{M. J. A. de Dood}
\affiliation{Huygens Kamerlingh-Onnes Laboratory, Leiden University, Niels Bohrweg 2, 2333 CA Leiden, The Netherlands}
\email{dood@physics.leidenuniv.nl}

\date{\today}

\begin{abstract}
  We optimize the design of NbN    nanowire superconducting single photon detectors, using the recently discovered position dependent detection efficiency in these devices. This   optimized design of  meandering wire NbN detectors maximizes absorption at positions where photon detection is most efficient   by altering the field distribution across the wire. In order to calculate the response of the detectors with different geometries, we use a monotonic local detection efficiency from a nanowire and optical absorption distribution via finite-different-time-domain simulations. The calculations predict a trade-off between average absorption and   absorption at the edge,   leading to a predicted optimal wire width close to 100 nm for 1550 nm wavelength, which drops to 50 nm wire width for 600 nm wavelength. The absorption at the edges can be enhanced by depositing a silicon nanowire on top of the superconducting nanowire, which improves both the total absorption efficiency as well as the internal detection efficiency of meandering wire structures.   The proposed structure can be integrated in a relatively simple cavity structure to reach absorption efficiencies of 97\% for perpendicular and 85\% for parallel polarization. 

\end{abstract}


\maketitle

\section{Introduction}
Thin and narrow superconducting nanowires to which a bias current is applied can be used to detect single photons \cite{Gol'tsman01,Hadfield09}. Absorption of a single photon of visible light is sufficient to switch part of the wire from the superconducting to the normal state and results in a measurable voltage. An appropriate design of the biasing electronics removes the current from the device once it is in the normal state, allowing the wire to self-reset to the superconducting state on a nanosecond timescale~\cite{Kerman06}. The specific benefits of these detectors are their broad spectral range (from visible to infrared wavelengths) \cite{Marsili11}, ultra-low dark-count rates \cite{Yamashita11}, excellent timing resolution \cite{Pearlman05}, and high detection efficiency \cite{Marsili13}. This makes these detectors very suitable for use in quantum optics \cite{Ikuta13}, quantum communication \cite{Sasaki11}, and life sciences \cite{Gemmell13}.

To optimize optical absorption in these superconducting single photon detectors (SSPDs), a meandering nanowire geometry embedded inside an optical cavity is preferred \cite{Rosfjord06,Marsili13,Miki09}. Device efficiencies close to 100\% can only be achieved when a bias current ($I_b$) is applied that creates a current density sufficiently close to the critical current density ($j_c$) of the superconducting material. This threshold for efficient detection depends on photon energy~\cite{Marsili11,Semenov05} and higher bias currents are needed to detect photons with lower energy (i.e. longer wavelength). An open challenge in this field is to extend the boundary for efficient detection of infrared radiation to the red. In general, thin and narrow wires are more difficult to fabricate but are expected to be better for detecting low energy photons because the absorption of a single photon in these devices affects a larger fraction of the total available Cooper pairs compared to a thicker and wider wire. However, thicker and wider wire are easier to fabricate with fewer defects~\cite{Gaudio14,Charaev16} and, more importantly, imply a lower kinetic inductance of the wire which yields detectors with a desirable faster response time~\cite{Kerman06}.   As a numerical example we estimate the kinetic inductance of a standard, 100 nm wide meandering wire covering a 10$\times$10~$\mu$m$^2$ area to be $L_k = \lambda_k L/A = 450$~nH. Here $A$ is the cross sectional area of the wire, $L$ is the total wire length and $\lambda_k \sim 360$ pH nm is the kinetic inductivity of NbN. For a 150 nm wide wire covering the same area, $L/A$ is reduced by a factor $(1.5)^2$ lowering the reset time from an estimated 9~ns to 4~ns.

A compromise between detection efficiency and speed needs to be found, which seems to be set by the material of the nanowire. We have reported a linear relation between a threshold current and photon energies for 150 nm wide NbN nanodetectors~\cite{Renema14}, i.e. significantly wider than the nanowires in NbN based SSPDs \cite{Gol'tsman01,Hadfield09}. In this simple geometry the device critical current is closer to the critical current density of the material~\cite{Charaev16}. For (meandering) wires the device critical current is limited by other factors such as current crowding in bends~\cite{Clem11}, fabrication defects along the wire~\cite{Hortensius12,Gaudio14}, thickness variations of the wire, or inhomogeneities intrinsic to the superconducting material~\cite{Hortensius13}.

Our recent experimental results~\cite{Renema15} as well as the improved understanding of the detection mechanism in SSPDs~\cite{Vodolazov15,Engel15,Engel15review} show that the edges of an SSPD are generally more efficient than the center of the wire. Hence, the internal detection efficiency ($IDE$) depends on position , photon energy   and bias current and is described by a quantity that we call the local detection efficiency $LDE(x, I_b)$.   For a given wavelength, uniform   detection efficiency is only reached for a bias current that exceeds the threshold current for all positions in the wire.   Numerical simulations show that   the variations in the local detection efficiency become more prominent for wider wires ($\ge$80 nm for 4 nm thick NbN film), lower bias currents and lower photon energy. The effects of a non-uniform $LDE$ will become noticeable in detectors that cannot be biased close to the material critical current density and in the roll-off regime of the efficiency as a function of bias-current. This regime becomes more important for infrared photon detection and in various SSPD based devices where multiple nanowires are operated in parallel to obtain photon number   resolving  ~\cite{Divochiy08,Marsili11PhtnRes} or a multi-pixel   detectors. Absorption of a photon by one of the nanowires creates a sudden increase in the resistance of that nanowire, causing the current to redistribute between the other nanowires and the input resistance of the amplifier. If the device is biased too close to the critical current the redistribution of current drives the other wires into a resistive state.

In this article, we present novel designs of SSPDs where the absorption   and internal detection   efficiency are optimized simultaneously.   These design considerations become important when the internal efficiency is less than 100\%. In this regime the $IDE$ depends on position and the total device $IDE$ can be optimized by tuning the electric field distribution.

\section{Local detection efficiency}

A known limitation of (meandering) wires is that the detection efficiency depends on the polarization of the incident light \cite{Anant08, Driessen09Eur}; at normal incidence absorption is more efficient when the electrical field vector of the light is aligned parallel to the wires, an effect governed by the boundary conditions on the edges of the nanowires. Recently, a design of a cavity-based structure was reported that tries to eliminate the polarization dependence and achieves up to 96\% absorption efficiency for both polarizations \cite{Zheng16}, assuming that the internal detection efficiency is uniform across the nanowire.

Detailed measurements of the polarization dependent response of both meandering wires \cite{Anant08} and nanodetectors \cite{Renema15} separate the optical absorption from the electronic detection process and reveal that the internal detection efficiency depends on polarization. This conundrum can be resolved if one assumes a position dependent local detection efficiency $LDE(x, I_b)$ in combination with a polarization and position dependent absorption $A_{\perp,\parallel}(x)$ of the detector. A microscopic detection model that uses quasiparticle diffusion and photon-assisted vortex entry \cite{Engel15,Renema14}, can explain the position dependence of the photon detection efficiency in term of an edge barrier for vortex entry. Within this model, absorption of a photon breaks Cooper pairs to create a cloud of quasi particles that diffuses through the wire. The dominant effect for the photon detection process is a reduction of the edge barrier for vortices. Photon detection occurs when a vortex crosses this energy barrier and travels across the wire and dissipates energy. When a photon is absorbed in the middle of the wire the barrier for vortex entry is affected to a much lesser extent than when a photon is absorbed at the edge of the wire, explaining the physical origin of more efficient detection at the edge of the wire.

We performed a study of a NbN nanodetector as a function of both wavelength and polarization of the incident light and employed detector tomography to separate the optical absorption efficiency ($\eta$) from the internal detection efficiency ($IDE$). A numerical calculation of the wavelength and polarization dependent absorption profile, together with the local detection efficiency $LDE(x, I_b)$ determines the response of the detector. We find~\cite{Renema15} that photon absorption events occurring at the edge of a 150~nm wide wire are much more likely to produce a detection event than those in the middle. Based on the experimental data this ``edge effect'' extends roughly 30~nm into the wire and forms the basis of an empirical model for the position dependent detection efficiency of a NbN detector.

This model can be summarized as follows. The internal detection efficiency $IDE(I_b)$ depends on the bias current $I_b$ and can be calculated from the absorption profile $A_{\perp,\parallel}(x)$ and the local detection efficiency $LDE(x,I_b)$
\begin{equation}
\label{eq:IDE}
IDE(I_b) = \frac{\int_{-w/2}^{w/2}{}A_{\perp,\parallel}(x) \cdot LDE(x,I_b)\mathrm{d}x}{\eta \cdot w},
\end{equation}
where $\eta$ is the average optical absorption over the wire
\begin{equation}
\label{eq:eta}
\eta_{\perp,\parallel} = \frac{1}{w} \int_{-w/2}^{w/2}A_{\perp,\parallel}(x)\mathrm{d}x.
\end{equation}
The internal response of an SSPD depends on photon energy and device bias current, and for a constant photon energy, the response increases exponentially as a function of bias current and saturates above a threshold current. Following Ref.~\cite{Renema15} we posit a relation between the local detection efficiency $LDE(x)$, bias current $I_b$ and threshold current $I_{th}(x)$, where the position dependence is expressed through the position dependence of the threshold current
\begin{equation}
\label{eq:LDE}
LDE(x,I_b)=
\begin{cases}
   \exp \left[\frac{\it(I_b-I_{th}(x))}{\it I^\ast}\right],& \text{if } I_b\leq I_{th}(x) \\
    1,              & \text{otherwise}
\end{cases}
\end{equation}
where $I^\ast$ = 0.65 $\mu$A is taken as a wavelength independent current scale that can be extracted from experiments by fitting the internal detection efficiency $IDE$ as a function of $I_{b}$ in Ref.~\cite{Renema15}. The function can be further specified through an assumption that the local threshold current $I_{th}(x)$ depends linearly on photon energy. This assumption is based on the observation that the position averaged response shows linear energy dependence~\cite{Renema14} and is further supported by numerical calculation of the microscopic detection model~\cite{Engel15,Renema15}.
\begin{equation}
I_{th}(x)=I_c-\gamma(x)E
\end{equation}
where $I_c$ is the device critical current and $E$ is the photon energy.

The absorption profile depends on the wavelength of the incident radiation and has to be obtained through numerical computation. The $\sim$~4 nm thick NbN film is much thinner than the skin depth of the relevant wavelength range. Hence, the absorption distribution is uniform over the thickness of the film and is a function of position $x$ across the nanowire only. We used a commercial finite-difference-time-domain method (FullWave package, RSoft \cite{RSoft}) and obtain the absorption profile for plane wave illumination at normal incidence as \cite{Jackson}
\begin{equation}
\label{eq:Abs}
A_{\perp,\parallel}(x) = \frac{P_{abs}(x)}{P_{total}} =\frac{\int^{t}_{0}\frac{1}{2}\omega \varepsilon_0 \rm Im\it(\varepsilon_{_{NbN}})|E_{\perp,\parallel}(x,y)^{\rm 2}|\it dy}{P_{total}},
\end{equation}
where $w$ is the width of the wire, $P_{total}$ is the power density of illumination incident on the wire, $P_{abs}(x)$ is the absorbed power density as a function of position $x$ across the wire width, $\omega$ is the angular frequency of the incident light, $\varepsilon_0$ is the vacuum permittivity, $t$ = 4.35 nm is the thickness of the NbN film obtained from ellipsometry measurements, and $|E_{\perp,\parallel}(x,y)^2|$ is the (polarization dependent) electric field intensity in the wire.

It should be noted that the model put forward here contains a number of approximations and assumptions to capture the strong decrease in detection efficiency in the center of the wire with a minimum number of parameters.   We use the exponential dependence of the $LDE$ given by Eq.~\ref{eq:LDE} in combination with a wavelength independent $I^\ast$ to capture the relevant details for NbN nanowires. The functional dependence of the $LDE$ is not universal and results on amorphous superconducting nanowires~\cite{Kozorezov15,Caloz16} suggest an error function in combination with a wavelength dependent $I^\ast$. We have verified that replacing the exponential function with an error function leads to small changes in the computed values,   but will not qualitatively change the observations and conclusions of our current work.

We use the simple model with fewer free parameters and define the response through the function $\gamma(x)$, which is the only unknown. This function can be obtained through a numerical inversion procedure where we use the value of $\gamma(x)$ sampled at 9 positions across the wire as fit parameters. Details of this procedure can be found in Ref.~\cite{Renema15}. The resulting function $\gamma(x)$ reflects the fact that the edges of the wire are more efficient at photon detection at reduced current compared to the center. The experimental data are consistent with highly efficient edges that extend $\sim$30-40 nm into the wire. Important additional support for this empirical model can be found by comparing the predictions of the model to the measured polarization-dependent absorption and detection efficiency of a set of meandering NbN SSPDs by Anant et al. \cite{Anant08}. These data show that the internal detection efficiency   for a telecommunication wavelengths of 1550 nm   is less than 100\% in devices and depends on polarization. Good quantitative agreement between these data and the empirical model can be achieved if one assumes that   these meandering wire devices cannot be biased beyond a current density that is $\sim$ 90\% of the material limited critical current density measured on the nanodetector   \cite{Charaev16,Renema15,Kerman06}

In this article we are interested in the implications of the local detection efficiency for future detector design and we need a stable and reliable implementation of the function $\gamma(x)$. The original empirical model shows small, non-monotonic, variations in detection efficiency closer to the edges of the wire that are currently not understood. The amplitude of these variations is comparable to the accuracy of the numerical inversion procedure but may affect the numerical optimization of the detector design. To avoid these numerical issues we use the original data   and procedures   from Ref.~\cite{Renema15}.   We adapt the   Tikhonov regularization procedure   that penalizes solutions where variations between adjacent points are large by a constrained optimization\cite{RenemaThesis} that demands the function $\gamma(x)$ to be  monotonic . A criterion that increases the value of $\chi^2$ by 20\% compared to the minimum value is sufficient to achieve this.

\begin{figure}[htb!]
\centering
\includegraphics[width=75mm]{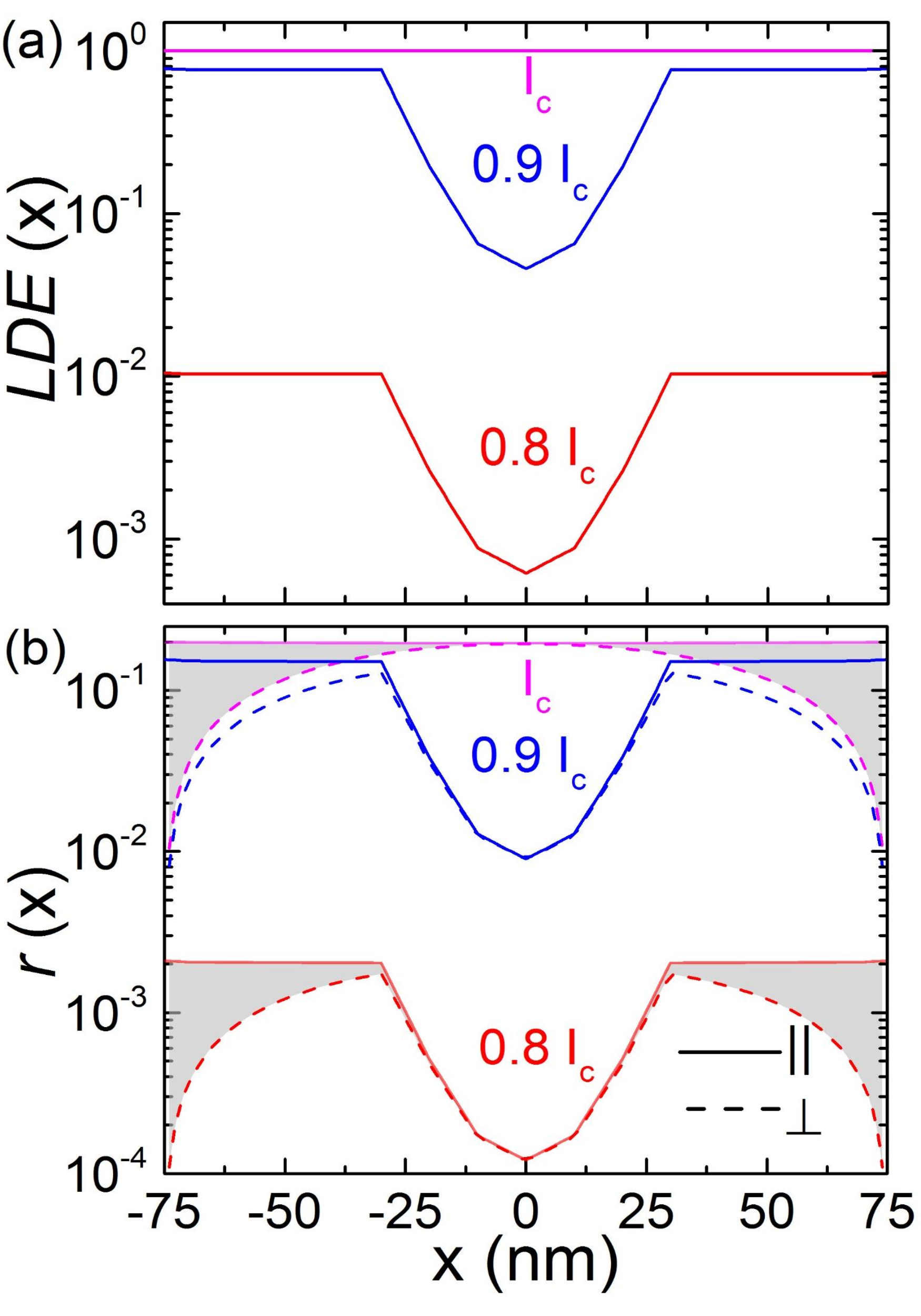}
 \caption{\label{fig:LDEIth}(a) Local detection efficiency across a 150 nm wide NbN wire at values of the bias current equal to 0.8$I_{_{C}}$, 0.9$I_{_{C}}$ and $I_{_{C}}$. The curves are determined from experimental data. (b) Detection response distribution $r(x)$ across the 150 nm wide NbN wire at the same bias currents as shown in (a). The solid and dashed curves represent parallel ($\|$) and perpendicular ($\bot$) polarization, respectively. The grey area between the solid and dashed curves at bias currents of 0.8$I_{_{C}}$ and $I_{_{C}}$ shows the difference of the local response between the two polarizations.}
 \end{figure}

Figure~\ref{fig:LDEIth}(a) shows the results of the $LDE(x)$ at bias currents $I_b$ = 0.8$I_{c}$, 0.9$I_{c}$ and $I_{c}$ for photons with a wavelength of 1550 nm (0.8 eV energy) using the monotonic version of $\gamma(x)$. As can be seen, the edges of the wire have much higher local detection efficiency compared to the center for lower bias currents. As the bias current increases, the detection efficiency at the edge starts to saturate and for currents close to the critical current of the nanodetector the efficiency $LDE(x)$ equals one over the entire width of the wire. Figure~\ref{fig:LDEIth}(b) shows the calculated local response $r(x)=A_{\perp,\parallel}(x)\cdot LDE(x,I_b)$ for both polarizations and values of the bias currents ($I_b$ = 0.8 $I_c$, 0.9 $I_c$ and $I_c$). For the highest bias current this local response distribution $r(x)$ is not limited by the $LDE(x)$ and thus reflects the profile of the absorption distribution $A_{\perp,\parallel}(x)$. The area between the solid and dashed curve for each bias current illustrates the difference in local response for the two polarizations. The total detector response that is accessible in the experiment as an observed count rate is proportional to the local response $r(x)$ integrated over position. We emphasize that this calculation depends on the geometry of the detector, the energy of the photons and the bias current through the nanowire.

\section{Results and discussion}

We compute the internal detection efficiency of differently designed SSPDs and visualize this by plotting the total detector response as a function of the optical absorption. Detectors that reach 100\% internal detection efficiency can always be designed for high energy photons and bias currents close to the material critical current density. Instead we are interested in the transition regime where the detectors become inefficient due to a combination of low bias current and/or low photon energy. In this regime the performance of the detectors can be improved by design of the SSPD and its surroundings. Such designs extend the sensitivity of SSPDs towards infrared wavelengths without compromising other device parameters such as response time and jitter. In practice, the regime of less than 100\% detection efficiency is reached for photon detection at 1550 nm with NbN meandering wire detectors that cannot be biased to the material critical current density due to fabrication imperfections and bends in the wire~\cite{Charaev16,Anant08,Renema15}.

\subsection{Optimal response of meandering wire detectors}

We first calculate the response of SSPDs with a constant wire width $w$ = 150 nm while varying the distance between the wires. We use finite-difference-time-domain (FDTD, FullWave package, RSoft \cite{RSoft}) simulations in two dimensions by approximating a meandering SSPD by an infinitely large array of wires and limit ourselves to a plane wave incident at normal incidence.  The 4.35 nm thick NbN wire ($n_{_{NbN}}= 5.23+5.82i$ at 1550 nm) is placed on a semi-infinite sapphire substrate ($n_{_{sapphire}}= 1.75$ at 1550 nm) and covered by a 2 nm thick oxide layer $\rm NbN_{x}O_{y}$ ($n_{_{NbNO}}= 2.28$, independent of wavelength). Calculations for other wavelengths take into account the small change in refractive index of the substrate~\cite{Dodge69} ($n_{_{sapphire}}= 1.77$  at 600 nm) as well as the important wavelength dependence of the refractive index of NbN. We use $n_{_{NbN}} = 1.98 + 3.22i$ at 600 nm wavelength and $n_{_{NbN}} = 3.51 + 4.32i$ at 1000 nm  wavelength~\cite{SahinThesis}.

\begin{figure}[htbp]
\centering
\includegraphics[width=75mm]{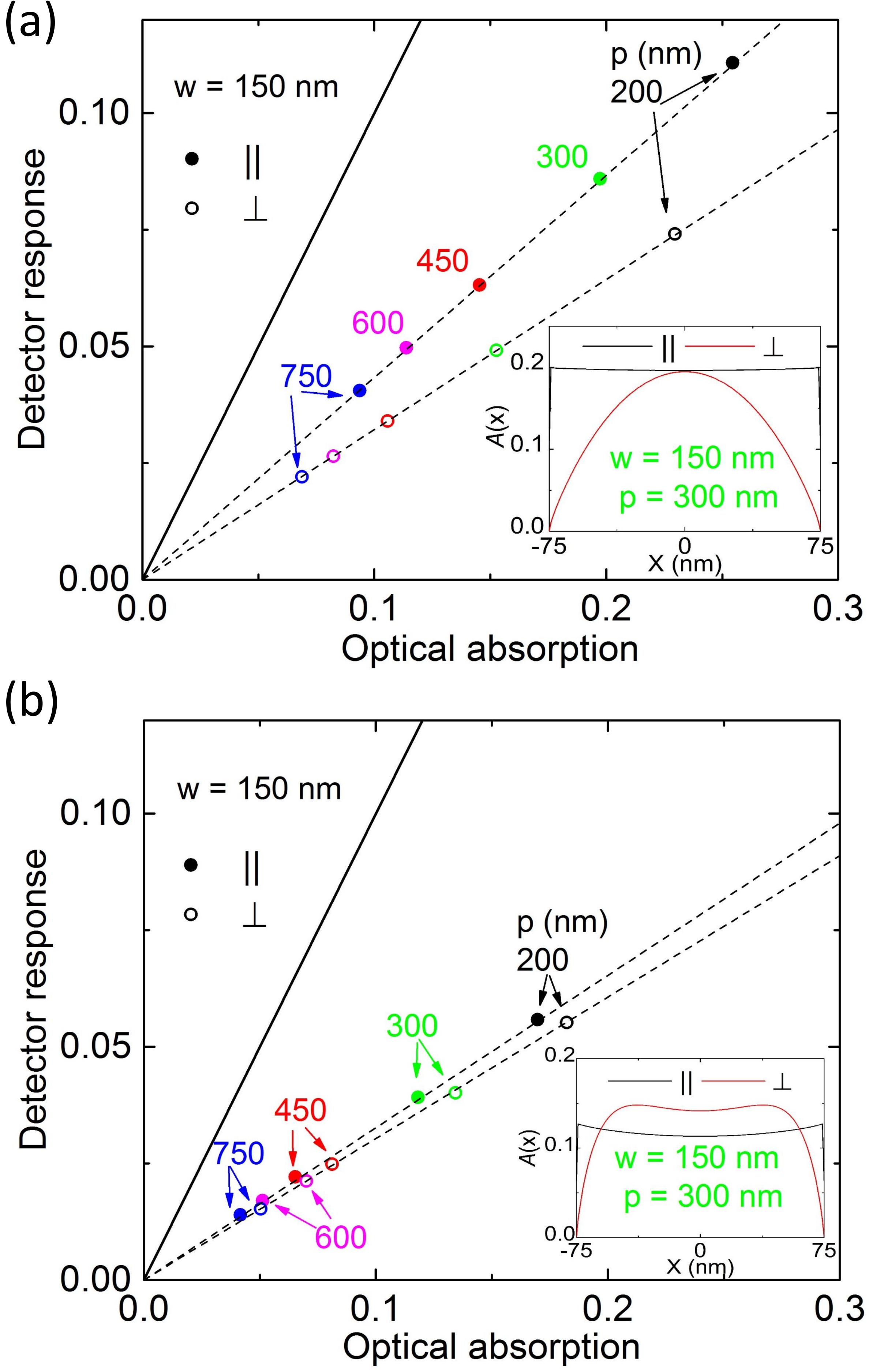}
\caption{\label{fig:ConstantWire}Calculated detector response as a function of optical absorption for meandering SSPDs with constant wire width ($w$ = 150 nm) and different pitches for wavelengths of 1550 nm (a) and 600 nm (b). The closed and open symbols represent parallel ($\|$) and perpendicular ($\bot$) polarization, respectively. The numbers next to the data points refer to the varying pitch $p$ expressed in nm. The two insets show a typical absorption distribution across the wire for structure with $w$ = 150 nm and $p$ = 300 nm.}
\end{figure}

Figure~\ref{fig:ConstantWire} shows the calculated detector response as a function of integrated optical absorption for different values of the pitch $p$ between the wires ($p$ = 200~nm, 300~nm, 450~nm, 600~nm and 750~nm). The solid line with slope one indicates a detector with 100\% internal detection efficiency for which each absorbed photon triggers the detector. Calculations are shown for wavelengths of 1550 nm (a) and 600 nm (b) for polarization parallel (closed symbols) and perpendicular to the wires (open symbols). The bias current $I_{b}$ is set to 0.9$I_{c}$ for a wavelength of 1550 nm, and 0.75$I_{c}$ for 600 nm. It should be noted that these two wavelengths and bias currents were chosen to compare a situation where the wire width $w$ is much smaller than the free space wavelength $w = \lambda/10$ to a situation where the wire width is comparable to the wavelength $w = \lambda/4$. The insets in Fig.~\ref{fig:ConstantWire} show the absorption distribution $A_{\perp,\parallel}(x)$ for the two different wavelengths and polarization of the incident light and show the almost constant absorption profile for polarized light with the $E$-field vector parallel to the wires. For perpendicularly polarized light absorption occurs more in the center of the wire and the beginning of an oscillation is observed for a wavelength of 600 nm.

As can be seen in the figure, the internal detection efficiency of meandering wires is proportional to the fill fraction $f$, defined as the ratio of wire width $w$ over the pitch $p$. Numerical calculations of the absorption profile of the various structures indeed show that the absorption distribution $A_{\perp,\parallel}(x)$ is nearly identical to that of an isolated wire, resulting in an average absorption $\eta$ that is proportional to $f$\cite{Driessen09Eur}. For a wavelength of 1550 nm the $IDE$ of the SSPDs (slope of the dashed line) with parallel polarization is higher than that of SSPDs with perpendicular polarization, which is explained by the difference in absorption profile $A_{\perp,\parallel}(x)$ for the two polarizations. For a wavelength of 600 nm wavelength, the average absorption $\eta$ of parallel polarization is lower than that of the perpendicular case, and compensates the lower response of the less efficient center part of the wire. The observed polarization dependence of the meandering wire is lower than what is expected based on the optical absorption alone.

\begin{figure}[htbp]
\centering
\includegraphics[width=75mm]{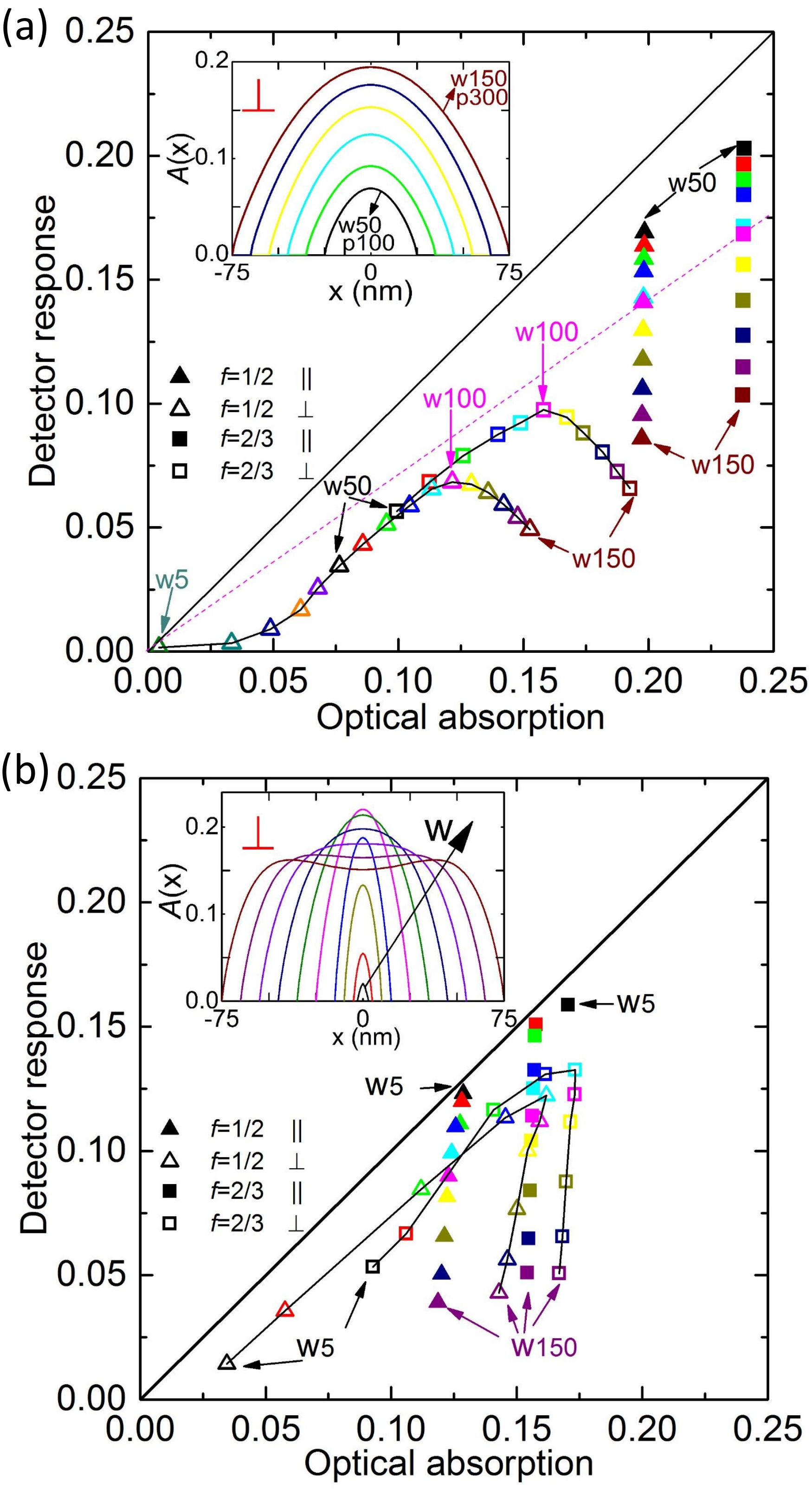}
\caption{\label{fig:ConstantFilling}Calculated detector response as a function of optical absorption for SSPD meandering structures with varying wire width and constant fill factor. Data are shown for a fill factor of 1$\slash$2 (triangles) and 2$\slash$3 (squares) and for wavelengths of 1550 nm (a) and 600 nm (b). The closed and open symbols refer to parallel ($\|$) and perpendicular ($\bot$) polarized light, respectively. The arrow indicates increasing width of the wire. The two insets show the absorption distribution for perpendicular polarized light as a function of wire width (fill factor 1$\slash$2).   Numbers next to the data points indicate the wire width in nm. For 1550 nm wavelength (a), the wire width is varied from 150 nm to 50 nm in steps of 10 nm. Additional points are calculated for 40, 30, 20, 10 and 5 nm width for $f$ = 1/2 and perpendicular polarization. For the 600 nm data (b), the width is varied from 150 nm to 10 nm in 20 nm steps, with additional points at 20 nm and 5 nm width.  }
\end{figure}

Figure~\ref{fig:ConstantFilling} shows the response of meandering SSPDs with constant fill factor as a function of optical absorption calculated for different wire widths. Data are shown for 1550 nm (a) and 600 nm wavelength (b) for parallel and perpendicular polarization (closed and open symbols). We compare structures with a fill factor $f = 1/2$ (triangles) and $f = 2/3$ (squares). To calculate the response of wires that are more narrow than 150 nm we omit the central part of the 150 nm wide $I_{th}(x)$ curve, leaving only the highly efficient edges of the wire. This procedure is motivated by numerical calculations of the $LDE(x)$ \cite{Engel15} that show that removing the central part of the curve of a wide wire correctly predicts the behavior of narrower wires. This procedure yields good quantitative agreement with the experimental data for different SSPDs with different width of the wires~\cite{Renema15}. We vary the wire width $w$ from 50 nm to 150 nm in steps of 10 nm. For perpendicular polarization and a fill factor $f = 1/2$ additional calculations are done for wire widths of 5 nm, 10 nm, 20 nm, 30 nm, 40 nm, 50 nm.

For parallel polarization (solid symbols), the absorption distribution across the wire is almost uniform and the average absorption $\eta$ is determined by the value of $f$ and the datapoints form a vertical line in the figure. The difference between the data points reflects the change in internal detection efficiency as a function of wire width $w$. The data for perpendicularly polarized light for a constant fill fraction $f$ and increasing $w$ (direction of the arrow in Fig.~\ref{fig:ConstantFilling})(a) show a maximum in detector response for a wire width around $w$ = 100 nm, for both values of the fill factor. The internal detection efficiency $IDE$ increases sharply with wire width for very narrow wires and decreases as $w$ increases beyond $w$ = 100 nm. This behavior originates from the dependence of the $LDE(x)$ as a function of wire width in combination with the non-uniform absorption distribution for perpendicularly polarized light.

With increasing wire width, the average of $LDE(x)$ decreases, but is compensated by an increase of the average absorption $\eta$, as shown in the inset of Fig.~\ref{fig:ConstantFilling}(a). Because the $LDE(x)$ is approximately constant for the first $\sim$ 30~nm from the edges (Fig.~\ref{fig:LDEIth}) and the average absorption increases with the wire width, the response of the detector increases as a function of wire width until $w\sim60$~nm. For a wire width beyond $\sim60$~nm, there is a trade-off for the wire to be narrow enough to have a high $LDE(x)$ and to be wide enough to have high optical absorption $A_{\perp,\parallel}(x)$. This trade-off depends on the detailed shape of the absorption, the bias current relative to the threshold current and the wavelength of the light. For the parameters considered in this article the trade-off results in an optimal value of $w \approx 100$~nm, almost three times the width of the edge effect in the $LDE(x)$ profile.

The calculations for a wavelength of 600~nm show a qualitatively different behavior with calculations done for nanowire widths set to 5~nm, 10~nm, 20~nm, 30~nm, 50~nm, 70~nm, 90~nm, 110~nm, 130~nm and 150~nm. Surprisingly the optimum wire width is only 50 nm and the internal efficiency shows a rapid rather than a gradual drop for wider wires. The inset in Fig.~\ref{fig:ConstantFilling}(b) shows that the optical absorption increases with width until the wire width reaches a value of 50 nm. For wider wires the absorption profile $A_{\perp,\parallel}(x)$ shows oscillations and the difference between the two polarizations disappears. For longer wavelengths there is a clear trade-off between higher absorption in wider wires and less efficient detection in the center of the wire. This trade-off no longer applies for a wavelength of 600 nm and the optimum internal efficiency is close to twice the width of the edge effect.

\subsection{Enhanced internal detection efficiency}

To take advantage of the higher efficiency at the edges of NbN nanowires, structures need to be designed that direct the absorption to these edges. Here we give an example of a dielectric structure that enhances edge absorption and enhances device performance. We limit ourselves to simple designs based on a patterned layer of silicon or GaAs on top and in between the nanowires so that the structure may be fabricated with current state-of-the-art nanotechnology.

\begin{figure}[htbp]
\centering
\includegraphics[width=75mm]{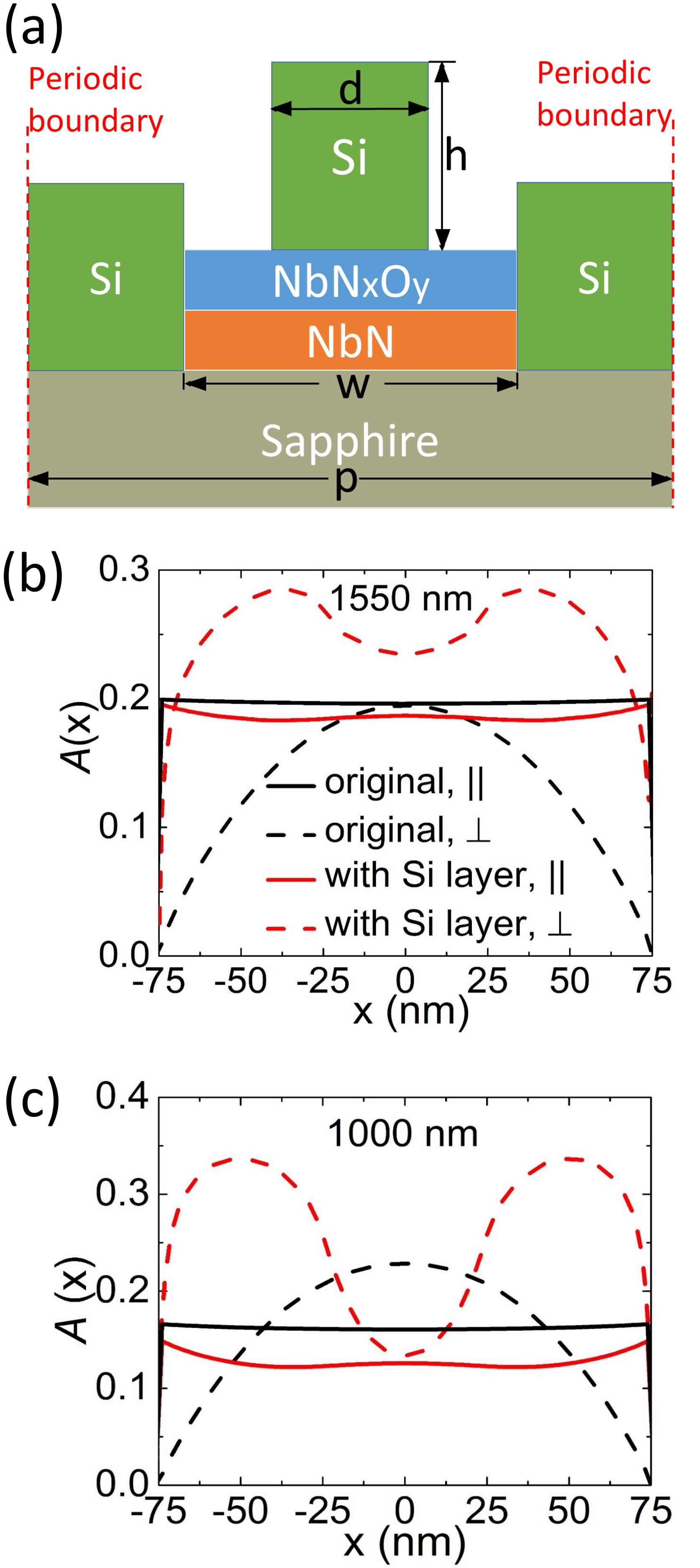}
\caption{\label{fig:SiLayer}(a) Si dielectric layer (n = 3.45 @ 1550 nm) used to enhance the absorption in perpendicular polarization at wavelength of 1550 nm. $w$ = 150 nm, $p$ = 300 nm, $d$ = 50 nm, $h$ = 30 nm. The periodic boundary is used to simulate the whole meander SSPD. (b)-(c) The absorption distribution across the NbN wire with (solid) and without (dash) Si layer.}
 \end{figure}

The suppressed absorption for perpendicular polarization at the edges is caused by the boundary conditions and the large mismatch between the dielectric constants of the vacuum and the NbN wire \cite{Zheng16}. To decrease the optical impedance mismatch and enhance the absorption, a dielectric layer between the vacuum and the NbN wire can be used. We take a NbN wire with $w$ = 150 nm and $p$ = 300 nm as a starting point to ensure that the wire is wider than the optimal width for 1550 nm. Figure~\ref{fig:SiLayer}(a) shows a coarsely optimized structure that enhances absorption at the edges and uses silicon (refractive index $n$ = 3.45 at 1550 nm wavelength) wires in between and on top of the NbN nanowire. Optimization of the structure for maximal absorption for perpendicular polarization at 1550 nm wavelength gives a close to optimum geometry of a Si layer of $h$ =  30 nm thickness shaped into $d$ = 50 nm wide Si wires on top of the 150 nm wide NbN nanowire.   We have constrained the optimization to conservative aspect rations of the silicon wires and set an equal thickness of the silicon in between and on top of the NbN wires to ensure that the design can be fabricated with state-of-the art technology. We note that a similar structure was fabricated for a different purpose \cite{Heath15}. The gap between the silicon wires enhances the field inside the gap and enhances optical absorption at the edges of the nanowire. We have explored less complicated   designs that cover the entire detector with a uniform layer,   and designs   that only have a nanowire on top of the NbN nanowire or in between the NbN nanowire.   These designs   reduce the impedance mismatch, but are not effective as a design that enhances the absorption at the edge of the wire.

Figure~\ref{fig:SiLayer}(b) shows the calculated absorption profile for the optimized structure and shows that the overall absorption is enhanced for perpendicular polarization. More importantly, the largest increase in absorption occurs at the edges for perpendicular polarization. We expect that the exact width and thickness of the Si structure are not very critical design parameters and that enhanced performance can be demonstrated for similar designs and for a wide wavelength range. To show the improved performance over a broad spectral range Fig.~\ref{fig:SiLayer}(c) shows the calculated absorption profile for 1000 nm wavelength using the geometric parameters of the structure optimized for 1550 nm wavelength. The refractive index of silicon changes to $n$ = 3.57 at 1000 nm wavelength~\cite{Green95}. For both wavelengths a minor change is observed in the absorption profile for parallel polarization, while the structure designed for 1550 nm is also effective in pushing the absorption towards the edges for 1000 nm wavelength.

Table~\ref{table1} contains a quantitative comparison of the average absorption $\eta$, the $IDE$ and the relative change in these numbers when comparing the original NbN meandering wire structure with the enhanced structure with the silicon wire. Results are included for both polarizations for detectors biased at the same bias current of 0.9 $I_c$. The larger energy of the 1000 nm photons causes the $IDE$ to increase by a factor 2 when comparing to the 1550 nm result, reflecting the linear exchange between photon energy and bias current. We are interested in the relative change in the detection efficiency $\eta$ and the $IDE$, which is small (less than 6\%) for parallel polarization at 1550 nm wavelength. This reflects the fact that the change in field distribution for parallel polarization is minimal. Comparing to the results for 1000 nm wavelength we see that the overall absorption $\eta$ is lowered by 20\% for parallel polarization while the $IDE$ is slightly decreased by 1.4\%. The field distribution is nearly uniform for parallel polarization and is, to a good approximation, independent of wavelength. The field distribution is significantly changed for perpendicular polarization and leads to a 70\% increase in the absorption accompanied by a 32\% increase in $IDE$.

\begin{figure}[htbp]
\centering
\includegraphics[width=75mm]{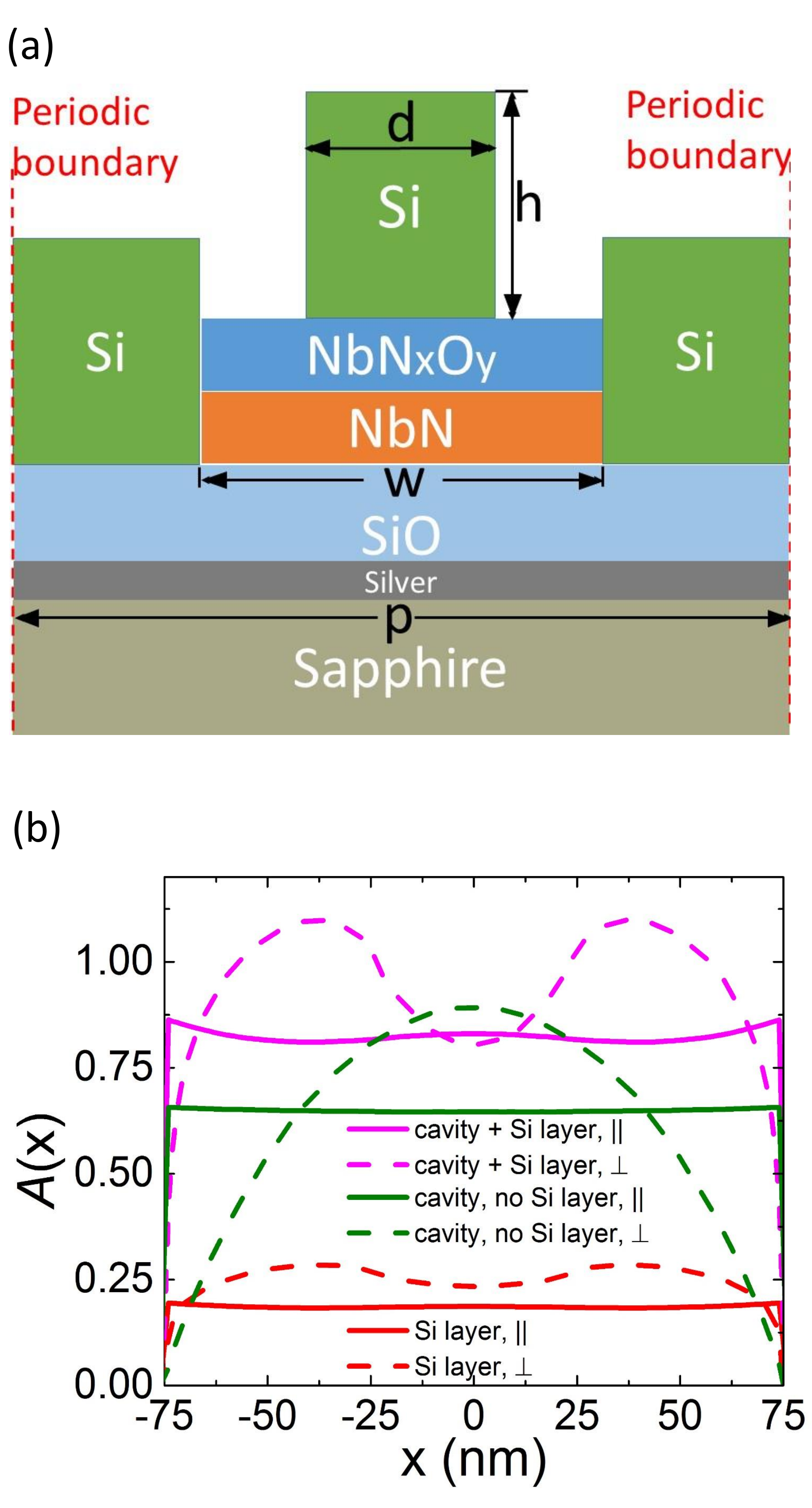}
\caption{\label{fig:Cavity} (a) Design of cavity structure based with enhanced edge absorption. The cavity structure is realized by inserting a 124 nm thick SiO layer ($n$ = 1.90  @ 1550 nm) and a 100 nm thick Ag later ($n$ = 0.51 + 10.71$i$ @ 1550 nm) underneath the detector. (b) Calculated absorption distribution comparing the design with a Si layer (red curves) to the optimal cavity design (purple curves). These distributions should be compared to the absorption distribution of a cavity structure without Si layers (green curves). }
\end{figure}

The optical absorption and device performance may be further enhanced if the design of the detector layer can be integrated inside an optical cavity. We calculated the absorption distribution for a wavelength of 1550 nm using a relatively simple design of a cavity with a single, 100 nm thick, Ag mirror layer ($n$ = 0.51 + 10.71$i$ @ 1550 nm) beneath a 124 nm thick SiO spacer layer ($n$ = 1.90  @ 1550 nm \cite{Hass54}). An interference between the direct reflection and the reflection from the mirror underneath the detector minimizes the reflection. Figure~\ref{fig:Cavity} shows the design of the cavity structure and calculated field distribution. The optimum thickness of the SiO layer minimizes reflection and maximizes absorption. As can be seen in Fig.~\ref{fig:Cavity}(b) the total absorption is significantly enhanced while the absorption distribution remains largely unaffected. For comparison the absorption distribution calculated for a cavity structure without the Si wire is shown as well and resembles the distribution of the structure of a bare NbN wire. This greatly simplifies the design process because the problem of optimizing the cavity from optimizing the field distribution in the nanowire is effectively decoupled. A calculation of the absorption of the nanowire integrated in the cavity structure shows that the absorption can be enhanced to 97\% for perpendicular and 85\% for parallel polarization.  

Interestingly, these numbers are remarkably robust against a change of wavelength indicating a spectrally broadband performance of the structure. Table~\ref{table1} shows that increase in absorption efficiency at 1000 nm is  66\% while the increase in $IDE$ amounts to 33\%. We hypothesize that this insensitivity to incident wavelength is due to the relatively small size of the silicon structure compared to the wavelength. The 50 nm width of the silicon wire is only $\lambda$/20 wide for the shorter wavelength of 1000 nm. A more realistic estimate of the size parameter relevant to optical scattering phenomena would be to take the optical path length of the circumference of the silicon structure as a measure, i.e. $x \approx (2d+2h) n/\lambda = 0.55$, where $n$ is the refractive index of silicon. As a result we do not expect higher order resonances to appear that could significantly alter the field distribution.

  We have repeated the calculations using a best matching sigmoidal function for the $LDE$ to verify that the numbers given in Table~\ref{table1} are relatively insensitive to the exact choice of the function describing the $LDE$. This best match is obtained by matching the slope and value of the function at the bias current where 50\% efficiency is reached. The calculated $IDE$ is higher for both polarizations and detector designs when using the sigmoidal function. The relative change in $IDE$ is unchanged for parallel polarization while the relative enhancement for perpendicular polarization is reduced to 23\% for 1550 nm and 21\% for 1000 nm. Further research is needed to better understand and quantify the function describing the $LDE$ of NbN nanowires and we remind the reader that the detection efficiency as a function of bias current for NbN detectors is not consistent with the sigmoidal function reported for amorphous superconductors \cite{Renema13}.

\begin{table}
\caption{\label{table1}The average absorption and detector response for a bare meandering wire compared to the enhanced structure.   Calculations are done for a meander with $w$~=150 nm wires and a pitch $p$ = 300 nm.   For both wavelengths the detector is biased at a current of 0.9 $I_c$.}

\begin{tabular}{l | c | c  c | c  c | c c }
    $\lambda$   &        & \multicolumn{2}{c|}{Original} & \multicolumn{2}{c|}{Enhanced} &  \multicolumn{2}{c}{$\Delta$} \\
                &        & $\perp$  & $\parallel$        & $\perp$  & $\parallel$        & $\perp$  & $\parallel$ \\
    \hline
    1550 nm     & $\eta$ & 0.15 & 0.20 & 0.26 & 0.19 & 70\% & -5.6\% \\
    1550 nm     & $IDE$  & 0.32 & 0.44 & 0.43 & 0.44 & 32\% & 0\% \\
    \hline
    1000 nm     & $\eta$ & 0.16 & 0.16 & 0.27 & 0.13 & 66\% & -20\% \\
    1000 nm     & $IDE$  & 0.61 & 0.74 & 0.81 & 0.73 & 33\% & -1.4\% \\
\end{tabular}
\end{table}

\section{Conclusions}

We have shown that the design of meandering wire SSPDs can be optimized for detection of a particular wavelength by optimizing the $IDE$. This is a direct consequence of the measured local detection efficiency~\cite{Renema15} that predicts that wider wires biased at lower currents have higher detection efficiency for photons absorbed at the edges. We calculate the average absorption efficiency as well as the $IDE$ by taking into account the electrical field distribution obtained from FDTD simulations and the local detection efficiency through a model of photon assisted vortex entry. The experimentally measured local detection efficiency predicts that photon detection is most efficient for photons absorbed within 30-40 nm from the edge of the wire.

The optimum width of a NbN nanowire that maximizes the detector response depends on wavelength and varies from 100 nm for 1550 nm wavelength to 50 nm wide wires for 600 nm wavelength. Absorption of wires that are much more narrow than the wavelength is inefficient for incident light that is polarized with the $E$-field perpendicular to the wire. The optimum width is a compromise between maximizing the total absorption while keeping the $IDE$ high by avoiding absorption in the center of the wire. For shorter wavelengths this balance shifts to a width that is approximately twice the size of the edge and the design maximizes the $IDE$ with the overall absorption being more or less independent of width for both polarizations for wider wires.

The $IDE$ can be enhanced significantly by placing a dielectric nanowire on top of the superconducting nanowire. We demonstrate a coarsely optimized structure using a 50 nm wide and 30 nm thick silicon wire on top of a 150 nm wide NbN nanowire. The wider NbN nanowire is easier to fabricate and reduces the detector response time by lowering the kinetic inductance compared to more narrow wires. For the optimal design both the absorption efficiency and the $IDE$ are increase by 70\% and 32\%, respectively. Calculations for 1000 nm wavelength show that the enhancement is a broadband effect. More complex designs, not necessarily limited to the use of dielectrics, can be envisaged that enhance device performance further or that resonantly enhance the $IDE$ for a particular wavelength or wavelength range~\cite{Wang15}.

\section*{Acknowledgement}

We thank Martin van Exter for scientific discussions and critical reading of the final manuscript. This work was funded by the Foundation for
Fundamental Research on Matter (FOM), which is financially supported by the Netherlands Organization for Scientific Research (NWO).


\begin{thebibliography}{40}%
\makeatletter
\providecommand \@ifxundefined [1]{%
 \@ifx{#1\undefined}
}%
\providecommand \@ifnum [1]{%
 \ifnum #1\expandafter \@firstoftwo
 \else \expandafter \@secondoftwo
 \fi
}%
\providecommand \@ifx [1]{%
 \ifx #1\expandafter \@firstoftwo
 \else \expandafter \@secondoftwo
 \fi
}%
\providecommand \natexlab [1]{#1}%
\providecommand \enquote  [1]{``#1''}%
\providecommand \bibnamefont  [1]{#1}%
\providecommand \bibfnamefont [1]{#1}%
\providecommand \citenamefont [1]{#1}%
\providecommand \href@noop [0]{\@secondoftwo}%
\providecommand \href [0]{\begingroup \@sanitize@url \@href}%
\providecommand \@href[1]{\@@startlink{#1}\@@href}%
\providecommand \@@href[1]{\endgroup#1\@@endlink}%
\providecommand \@sanitize@url [0]{\catcode `\\12\catcode `\$12\catcode
  `\&12\catcode `\#12\catcode `\^12\catcode `\_12\catcode `\%12\relax}%
\providecommand \@@startlink[1]{}%
\providecommand \@@endlink[0]{}%
\providecommand \url  [0]{\begingroup\@sanitize@url \@url }%
\providecommand \@url [1]{\endgroup\@href {#1}{\urlprefix }}%
\providecommand \urlprefix  [0]{URL }%
\providecommand \Eprint [0]{\href }%
\providecommand \doibase [0]{http://dx.doi.org/}%
\providecommand \selectlanguage [0]{\@gobble}%
\providecommand \bibinfo  [0]{\@secondoftwo}%
\providecommand \bibfield  [0]{\@secondoftwo}%
\providecommand \translation [1]{[#1]}%
\providecommand \BibitemOpen [0]{}%
\providecommand \bibitemStop [0]{}%
\providecommand \bibitemNoStop [0]{.\EOS\space}%
\providecommand \EOS [0]{\spacefactor3000\relax}%
\providecommand \BibitemShut  [1]{\csname bibitem#1\endcsname}%
\let\auto@bib@innerbib\@empty
\bibitem [{\citenamefont {Gol'tsman}\ \emph {et~al.}(2001)\citenamefont
  {Gol'tsman}, \citenamefont {Okunev}, \citenamefont {Chulkova}, \citenamefont
  {Lipatov}, \citenamefont {Semenov}, \citenamefont {Smirnov}, \citenamefont
  {Voronov}, \citenamefont {Dzardanov}, \citenamefont {Williams},\ and\
  \citenamefont {Sobolewski}}]{Gol'tsman01}%
  \BibitemOpen
  \bibfield  {author} {\bibinfo {author} {\bibfnamefont {G.~N.}\ \bibnamefont
  {Gol'tsman}}, \bibinfo {author} {\bibfnamefont {O.}~\bibnamefont {Okunev}},
  \bibinfo {author} {\bibfnamefont {G.}~\bibnamefont {Chulkova}}, \bibinfo
  {author} {\bibfnamefont {A.}~\bibnamefont {Lipatov}}, \bibinfo {author}
  {\bibfnamefont {A.}~\bibnamefont {Semenov}}, \bibinfo {author} {\bibfnamefont
  {K.}~\bibnamefont {Smirnov}}, \bibinfo {author} {\bibfnamefont
  {B.}~\bibnamefont {Voronov}}, \bibinfo {author} {\bibfnamefont
  {A.}~\bibnamefont {Dzardanov}}, \bibinfo {author} {\bibfnamefont
  {C.}~\bibnamefont {Williams}}, \ and\ \bibinfo {author} {\bibfnamefont
  {R.}~\bibnamefont {Sobolewski}},\ }\href@noop {} {\bibfield  {journal}
  {\bibinfo  {journal} {Appl. Phys. Lett.}\ }\textbf {\bibinfo {volume} {79}},\
  \bibinfo {pages} {705} (\bibinfo {year} {2001})}\BibitemShut {NoStop}%
\bibitem [{\citenamefont {Hadfield}(2009)}]{Hadfield09}%
  \BibitemOpen
  \bibfield  {author} {\bibinfo {author} {\bibfnamefont {R.~H.}\ \bibnamefont
  {Hadfield}},\ }\href@noop {} {\bibfield  {journal} {\bibinfo  {journal} {Nat.
  Photonics}\ }\textbf {\bibinfo {volume} {3}},\ \bibinfo {pages} {696}
  (\bibinfo {year} {2009})}\BibitemShut {NoStop}%
\bibitem [{\citenamefont {Kerman}\ \emph {et~al.}(2006)\citenamefont {Kerman},
  \citenamefont {Dauler}, \citenamefont {Keicher}, \citenamefont {Yang},
  \citenamefont {Berggren}, \citenamefont {Gol'tsman},\ and\ \citenamefont
  {Voronov}}]{Kerman06}%
  \BibitemOpen
  \bibfield  {author} {\bibinfo {author} {\bibfnamefont {A.~J.}\ \bibnamefont
  {Kerman}}, \bibinfo {author} {\bibfnamefont {E.~A.}\ \bibnamefont {Dauler}},
  \bibinfo {author} {\bibfnamefont {W.~E.}\ \bibnamefont {Keicher}}, \bibinfo
  {author} {\bibfnamefont {J.~K.~W.}\ \bibnamefont {Yang}}, \bibinfo {author}
  {\bibfnamefont {K.~K.}\ \bibnamefont {Berggren}}, \bibinfo {author}
  {\bibfnamefont {G.}~\bibnamefont {Gol'tsman}}, \ and\ \bibinfo {author}
  {\bibfnamefont {B.}~\bibnamefont {Voronov}},\ }\href@noop {} {\bibfield
  {journal} {\bibinfo  {journal} {Appl. Phys. Lett.}\ }\textbf {\bibinfo
  {volume} {88}},\ \bibinfo {pages} {111116} (\bibinfo {year}
  {2006})}\BibitemShut {NoStop}%
\bibitem [{\citenamefont {Marsili}\ \emph
  {et~al.}(2011{\natexlab{a}})\citenamefont {Marsili}, \citenamefont {Najafi},
  \citenamefont {Dauler}, \citenamefont {Bellei}, \citenamefont {Hu},
  \citenamefont {Csete}, \citenamefont {Molnar},\ and\ \citenamefont
  {Berggren}}]{Marsili11}%
  \BibitemOpen
  \bibfield  {author} {\bibinfo {author} {\bibfnamefont {F.}~\bibnamefont
  {Marsili}}, \bibinfo {author} {\bibfnamefont {F.}~\bibnamefont {Najafi}},
  \bibinfo {author} {\bibfnamefont {E.}~\bibnamefont {Dauler}}, \bibinfo
  {author} {\bibfnamefont {F.}~\bibnamefont {Bellei}}, \bibinfo {author}
  {\bibfnamefont {X.}~\bibnamefont {Hu}}, \bibinfo {author} {\bibfnamefont
  {M.}~\bibnamefont {Csete}}, \bibinfo {author} {\bibfnamefont {R.~J.}\
  \bibnamefont {Molnar}}, \ and\ \bibinfo {author} {\bibfnamefont {K.~K.}\
  \bibnamefont {Berggren}},\ }\href@noop {} {\bibfield  {journal} {\bibinfo
  {journal} {Nano Lett.}\ }\textbf {\bibinfo {volume} {11}},\ \bibinfo {pages}
  {2048} (\bibinfo {year} {2011}{\natexlab{a}})}\BibitemShut {NoStop}%
\bibitem [{\citenamefont {Yamashita}\ \emph {et~al.}(2011)\citenamefont
  {Yamashita}, \citenamefont {Miki}, \citenamefont {Makise}, \citenamefont
  {Qiu}, \citenamefont {Terai}, \citenamefont {Fujiwara}, \citenamefont
  {Sasaki},\ and\ \citenamefont {Wang}}]{Yamashita11}%
  \BibitemOpen
  \bibfield  {author} {\bibinfo {author} {\bibfnamefont {T.}~\bibnamefont
  {Yamashita}}, \bibinfo {author} {\bibfnamefont {S.}~\bibnamefont {Miki}},
  \bibinfo {author} {\bibfnamefont {K.}~\bibnamefont {Makise}}, \bibinfo
  {author} {\bibfnamefont {W.}~\bibnamefont {Qiu}}, \bibinfo {author}
  {\bibfnamefont {H.}~\bibnamefont {Terai}}, \bibinfo {author} {\bibfnamefont
  {M.}~\bibnamefont {Fujiwara}}, \bibinfo {author} {\bibfnamefont
  {M.}~\bibnamefont {Sasaki}}, \ and\ \bibinfo {author} {\bibfnamefont
  {Z.}~\bibnamefont {Wang}},\ }\href@noop {} {\bibfield  {journal} {\bibinfo
  {journal} {Appl. Phys. Lett.}\ }\textbf {\bibinfo {volume} {99}},\ \bibinfo
  {pages} {161105} (\bibinfo {year} {2011})}\BibitemShut {NoStop}%
\bibitem [{\citenamefont {Pearlman}\ \emph {et~al.}(2005)\citenamefont
  {Pearlman}, \citenamefont {Cros}, \citenamefont {Slysz}, \citenamefont
  {Zhang}, \citenamefont {Verevkin}, \citenamefont {Currie}, \citenamefont
  {Korneev}, \citenamefont {Kouminov}, \citenamefont {Smirnov}, \citenamefont
  {Voronov}, \citenamefont {Gol'tsman},\ and\ \citenamefont
  {Sobolewski}}]{Pearlman05}%
  \BibitemOpen
  \bibfield  {author} {\bibinfo {author} {\bibfnamefont {A.}~\bibnamefont
  {Pearlman}}, \bibinfo {author} {\bibfnamefont {A.}~\bibnamefont {Cros}},
  \bibinfo {author} {\bibfnamefont {W.}~\bibnamefont {Slysz}}, \bibinfo
  {author} {\bibfnamefont {J.}~\bibnamefont {Zhang}}, \bibinfo {author}
  {\bibfnamefont {A.}~\bibnamefont {Verevkin}}, \bibinfo {author}
  {\bibfnamefont {M.}~\bibnamefont {Currie}}, \bibinfo {author} {\bibfnamefont
  {A.}~\bibnamefont {Korneev}}, \bibinfo {author} {\bibfnamefont
  {P.}~\bibnamefont {Kouminov}}, \bibinfo {author} {\bibfnamefont
  {K.}~\bibnamefont {Smirnov}}, \bibinfo {author} {\bibfnamefont
  {B.}~\bibnamefont {Voronov}}, \bibinfo {author} {\bibfnamefont
  {G.}~\bibnamefont {Gol'tsman}}, \ and\ \bibinfo {author} {\bibfnamefont
  {R.}~\bibnamefont {Sobolewski}},\ }\href@noop {} {\bibfield  {journal}
  {\bibinfo  {journal} {IEEE Trans. Appl. Supercond.}\ }\textbf {\bibinfo
  {volume} {15}},\ \bibinfo {pages} {579} (\bibinfo {year} {2005})}\BibitemShut
  {NoStop}%
\bibitem [{\citenamefont {Marsili}\ \emph {et~al.}(2013)\citenamefont
  {Marsili}, \citenamefont {Verma}, \citenamefont {Stern}, \citenamefont
  {Harrington}, \citenamefont {Lita}, \citenamefont {Gerrits}, \citenamefont
  {Vayshenker}, \citenamefont {Baek}, \citenamefont {Shaw}, \citenamefont
  {Mirin},\ and\ \citenamefont {Nam}}]{Marsili13}%
  \BibitemOpen
  \bibfield  {author} {\bibinfo {author} {\bibfnamefont {F.}~\bibnamefont
  {Marsili}}, \bibinfo {author} {\bibfnamefont {V.~B.}\ \bibnamefont {Verma}},
  \bibinfo {author} {\bibfnamefont {J.~A.}\ \bibnamefont {Stern}}, \bibinfo
  {author} {\bibfnamefont {S.}~\bibnamefont {Harrington}}, \bibinfo {author}
  {\bibfnamefont {A.~E.}\ \bibnamefont {Lita}}, \bibinfo {author}
  {\bibfnamefont {T.}~\bibnamefont {Gerrits}}, \bibinfo {author} {\bibfnamefont
  {I.}~\bibnamefont {Vayshenker}}, \bibinfo {author} {\bibfnamefont
  {B.}~\bibnamefont {Baek}}, \bibinfo {author} {\bibfnamefont {M.~D.}\
  \bibnamefont {Shaw}}, \bibinfo {author} {\bibfnamefont {R.~P.}\ \bibnamefont
  {Mirin}}, \ and\ \bibinfo {author} {\bibfnamefont {S.~W.}\ \bibnamefont
  {Nam}},\ }\href@noop {} {\bibfield  {journal} {\bibinfo  {journal} {Nat.
  Photonics}\ }\textbf {\bibinfo {volume} {7}},\ \bibinfo {pages} {210}
  (\bibinfo {year} {2013})}\BibitemShut {NoStop}%
\bibitem [{\citenamefont {Ikuta}\ \emph {et~al.}(2013)\citenamefont {Ikuta},
  \citenamefont {Kato}, \citenamefont {Kusaka}, \citenamefont {Miki},
  \citenamefont {Yamashita}, \citenamefont {Terai}, \citenamefont {Fujiwara},
  \citenamefont {Yamamoto}, \citenamefont {Koashi}, \citenamefont {Sasaki},
  \citenamefont {Wang},\ and\ \citenamefont {Imoto}}]{Ikuta13}%
  \BibitemOpen
  \bibfield  {author} {\bibinfo {author} {\bibfnamefont {R.}~\bibnamefont
  {Ikuta}}, \bibinfo {author} {\bibfnamefont {H.}~\bibnamefont {Kato}},
  \bibinfo {author} {\bibfnamefont {Y.}~\bibnamefont {Kusaka}}, \bibinfo
  {author} {\bibfnamefont {S.}~\bibnamefont {Miki}}, \bibinfo {author}
  {\bibfnamefont {T.}~\bibnamefont {Yamashita}}, \bibinfo {author}
  {\bibfnamefont {H.}~\bibnamefont {Terai}}, \bibinfo {author} {\bibfnamefont
  {M.}~\bibnamefont {Fujiwara}}, \bibinfo {author} {\bibfnamefont
  {T.}~\bibnamefont {Yamamoto}}, \bibinfo {author} {\bibfnamefont
  {M.}~\bibnamefont {Koashi}}, \bibinfo {author} {\bibfnamefont
  {M.}~\bibnamefont {Sasaki}}, \bibinfo {author} {\bibfnamefont
  {Z.}~\bibnamefont {Wang}}, \ and\ \bibinfo {author} {\bibfnamefont
  {N.}~\bibnamefont {Imoto}},\ }\href@noop {} {\bibfield  {journal} {\bibinfo
  {journal} {Phys. Rev. A}\ }\textbf {\bibinfo {volume} {87}},\ \bibinfo
  {pages} {010301(R)} (\bibinfo {year} {2013})}\BibitemShut {NoStop}%
\bibitem [{\citenamefont {Sasaki}\ \emph {et~al.}(2011)\citenamefont {Sasaki},
  \citenamefont {Fujiwara}, \citenamefont {Ishizuka}, \citenamefont {Klaus},
  \citenamefont {Wakui}, \citenamefont {Takeoka}, \citenamefont {Miki},
  \citenamefont {Yamashita}, \citenamefont {Wang}, \citenamefont {Tanaka},
  \citenamefont {Yoshino}, \citenamefont {Nambu}, \citenamefont {Takahashi},
  \citenamefont {Tajima}, \citenamefont {Tomita}, \citenamefont {Domeki},
  \citenamefont {Hasegawa}, \citenamefont {Sakai}, \citenamefont {Kobayashi},
  \citenamefont {Asai}, \citenamefont {Shimizu}, \citenamefont {Tokura},
  \citenamefont {Tsurumaru}, \citenamefont {Matsui}, \citenamefont {Honjo},
  \citenamefont {Tamaki}, \citenamefont {Takesue}, \citenamefont {Tokura},
  \citenamefont {Dynes}, \citenamefont {Dixon}, \citenamefont {Sharpe},
  \citenamefont {Yuan}, \citenamefont {Shields}, \citenamefont {Uchikoga},
  \citenamefont {Legre}, \citenamefont {Robyr}, \citenamefont {Trinkler},
  \citenamefont {Monat}, \citenamefont {Page}, \citenamefont {Ribordy},
  \citenamefont {Poppe}, \citenamefont {Allacher}, \citenamefont {Maurhart},
  \citenamefont {Langer}, \citenamefont {Peev},\ and\ \citenamefont
  {Zeilinger}}]{Sasaki11}%
  \BibitemOpen
  \bibfield  {author} {\bibinfo {author} {\bibfnamefont {M.}~\bibnamefont
  {Sasaki}}, \bibinfo {author} {\bibfnamefont {M.}~\bibnamefont {Fujiwara}},
  \bibinfo {author} {\bibfnamefont {H.}~\bibnamefont {Ishizuka}}, \bibinfo
  {author} {\bibfnamefont {W.}~\bibnamefont {Klaus}}, \bibinfo {author}
  {\bibfnamefont {K.}~\bibnamefont {Wakui}}, \bibinfo {author} {\bibfnamefont
  {M.}~\bibnamefont {Takeoka}}, \bibinfo {author} {\bibfnamefont
  {S.}~\bibnamefont {Miki}}, \bibinfo {author} {\bibfnamefont {T.}~\bibnamefont
  {Yamashita}}, \bibinfo {author} {\bibfnamefont {Z.}~\bibnamefont {Wang}},
  \bibinfo {author} {\bibfnamefont {A.}~\bibnamefont {Tanaka}}, \bibinfo
  {author} {\bibfnamefont {K.}~\bibnamefont {Yoshino}}, \bibinfo {author}
  {\bibfnamefont {Y.}~\bibnamefont {Nambu}}, \bibinfo {author} {\bibfnamefont
  {S.}~\bibnamefont {Takahashi}}, \bibinfo {author} {\bibfnamefont
  {A.}~\bibnamefont {Tajima}}, \bibinfo {author} {\bibfnamefont
  {A.}~\bibnamefont {Tomita}}, \bibinfo {author} {\bibfnamefont
  {T.}~\bibnamefont {Domeki}}, \bibinfo {author} {\bibfnamefont
  {T.}~\bibnamefont {Hasegawa}}, \bibinfo {author} {\bibfnamefont
  {Y.}~\bibnamefont {Sakai}}, \bibinfo {author} {\bibfnamefont
  {H.}~\bibnamefont {Kobayashi}}, \bibinfo {author} {\bibfnamefont
  {T.}~\bibnamefont {Asai}}, \bibinfo {author} {\bibfnamefont {K.}~\bibnamefont
  {Shimizu}}, \bibinfo {author} {\bibfnamefont {T.}~\bibnamefont {Tokura}},
  \bibinfo {author} {\bibfnamefont {T.}~\bibnamefont {Tsurumaru}}, \bibinfo
  {author} {\bibfnamefont {M.}~\bibnamefont {Matsui}}, \bibinfo {author}
  {\bibfnamefont {T.}~\bibnamefont {Honjo}}, \bibinfo {author} {\bibfnamefont
  {K.}~\bibnamefont {Tamaki}}, \bibinfo {author} {\bibfnamefont
  {H.}~\bibnamefont {Takesue}}, \bibinfo {author} {\bibfnamefont
  {Y.}~\bibnamefont {Tokura}}, \bibinfo {author} {\bibfnamefont {J.~F.}\
  \bibnamefont {Dynes}}, \bibinfo {author} {\bibfnamefont {A.~R.}\ \bibnamefont
  {Dixon}}, \bibinfo {author} {\bibfnamefont {A.~W.}\ \bibnamefont {Sharpe}},
  \bibinfo {author} {\bibfnamefont {Z.~L.}\ \bibnamefont {Yuan}}, \bibinfo
  {author} {\bibfnamefont {A.~J.}\ \bibnamefont {Shields}}, \bibinfo {author}
  {\bibfnamefont {S.}~\bibnamefont {Uchikoga}}, \bibinfo {author}
  {\bibfnamefont {M.}~\bibnamefont {Legre}}, \bibinfo {author} {\bibfnamefont
  {S.}~\bibnamefont {Robyr}}, \bibinfo {author} {\bibfnamefont
  {P.}~\bibnamefont {Trinkler}}, \bibinfo {author} {\bibfnamefont
  {L.}~\bibnamefont {Monat}}, \bibinfo {author} {\bibfnamefont {J.-B.}\
  \bibnamefont {Page}}, \bibinfo {author} {\bibfnamefont {G.}~\bibnamefont
  {Ribordy}}, \bibinfo {author} {\bibfnamefont {A.}~\bibnamefont {Poppe}},
  \bibinfo {author} {\bibfnamefont {A.}~\bibnamefont {Allacher}}, \bibinfo
  {author} {\bibfnamefont {O.}~\bibnamefont {Maurhart}}, \bibinfo {author}
  {\bibfnamefont {T.}~\bibnamefont {Langer}}, \bibinfo {author} {\bibfnamefont
  {M.}~\bibnamefont {Peev}}, \ and\ \bibinfo {author} {\bibfnamefont
  {A.}~\bibnamefont {Zeilinger}},\ }\href@noop {} {\bibfield  {journal}
  {\bibinfo  {journal} {Opt. Express}\ }\textbf {\bibinfo {volume} {19}},\
  \bibinfo {pages} {10387} (\bibinfo {year} {2011})}\BibitemShut {NoStop}%
\bibitem [{\citenamefont {Gemmell}\ \emph {et~al.}(2013)\citenamefont
  {Gemmell}, \citenamefont {McCarthy}, \citenamefont {Liu}, \citenamefont
  {Tanner}, \citenamefont {Dorenbos}, \citenamefont {Zwiller}, \citenamefont
  {Patterson}, \citenamefont {Buller}, \citenamefont {Wilson},\ and\
  \citenamefont {Hadfield}}]{Gemmell13}%
  \BibitemOpen
  \bibfield  {author} {\bibinfo {author} {\bibfnamefont {N.}~\bibnamefont
  {Gemmell}}, \bibinfo {author} {\bibfnamefont {A.}~\bibnamefont {McCarthy}},
  \bibinfo {author} {\bibfnamefont {B.}~\bibnamefont {Liu}}, \bibinfo {author}
  {\bibfnamefont {M.}~\bibnamefont {Tanner}}, \bibinfo {author} {\bibfnamefont
  {S.}~\bibnamefont {Dorenbos}}, \bibinfo {author} {\bibfnamefont
  {V.}~\bibnamefont {Zwiller}}, \bibinfo {author} {\bibfnamefont
  {M.}~\bibnamefont {Patterson}}, \bibinfo {author} {\bibfnamefont
  {G.}~\bibnamefont {Buller}}, \bibinfo {author} {\bibfnamefont
  {B.}~\bibnamefont {Wilson}}, \ and\ \bibinfo {author} {\bibfnamefont {R.~H.}\
  \bibnamefont {Hadfield}},\ }\href@noop {} {\bibfield  {journal} {\bibinfo
  {journal} {Opt. Express}\ }\textbf {\bibinfo {volume} {21}},\ \bibinfo
  {pages} {5005} (\bibinfo {year} {2013})}\BibitemShut {NoStop}%
\bibitem [{\citenamefont {Rosfjord}\ \emph {et~al.}(2006)\citenamefont
  {Rosfjord}, \citenamefont {Yang}, \citenamefont {Dauler}, \citenamefont
  {Kerman}, \citenamefont {Anant}, \citenamefont {Voronov}, \citenamefont
  {Gol'tsman},\ and\ \citenamefont {Berggren}}]{Rosfjord06}%
  \BibitemOpen
  \bibfield  {author} {\bibinfo {author} {\bibfnamefont {K.~M.}\ \bibnamefont
  {Rosfjord}}, \bibinfo {author} {\bibfnamefont {J.~K.~W.}\ \bibnamefont
  {Yang}}, \bibinfo {author} {\bibfnamefont {E.~A.}\ \bibnamefont {Dauler}},
  \bibinfo {author} {\bibfnamefont {A.~J.}\ \bibnamefont {Kerman}}, \bibinfo
  {author} {\bibfnamefont {V.}~\bibnamefont {Anant}}, \bibinfo {author}
  {\bibfnamefont {B.~M.}\ \bibnamefont {Voronov}}, \bibinfo {author}
  {\bibfnamefont {G.~N.}\ \bibnamefont {Gol'tsman}}, \ and\ \bibinfo {author}
  {\bibfnamefont {K.~K.}\ \bibnamefont {Berggren}},\ }\href@noop {} {\bibfield
  {journal} {\bibinfo  {journal} {Opt. Express}\ }\textbf {\bibinfo {volume}
  {14}},\ \bibinfo {pages} {527} (\bibinfo {year} {2006})}\BibitemShut
  {NoStop}%
\bibitem [{\citenamefont {Miki}\ \emph {et~al.}(2009)\citenamefont {Miki},
  \citenamefont {Takeda}, \citenamefont {Fujiwara}, \citenamefont {Sasaki},\
  and\ \citenamefont {Wang}}]{Miki09}%
  \BibitemOpen
  \bibfield  {author} {\bibinfo {author} {\bibfnamefont {S.}~\bibnamefont
  {Miki}}, \bibinfo {author} {\bibfnamefont {M.}~\bibnamefont {Takeda}},
  \bibinfo {author} {\bibfnamefont {M.}~\bibnamefont {Fujiwara}}, \bibinfo
  {author} {\bibfnamefont {M.}~\bibnamefont {Sasaki}}, \ and\ \bibinfo {author}
  {\bibfnamefont {Z.}~\bibnamefont {Wang}},\ }\href@noop {} {\bibfield
  {journal} {\bibinfo  {journal} {Opt. Express}\ }\textbf {\bibinfo {volume}
  {17}},\ \bibinfo {pages} {23557} (\bibinfo {year} {2009})}\BibitemShut
  {NoStop}%
\bibitem [{\citenamefont {Semenov}\ \emph {et~al.}(2005)\citenamefont
  {Semenov}, \citenamefont {Engel}, \citenamefont {Hubers}, \citenamefont
  {Il'in},\ and\ \citenamefont {Siegel}}]{Semenov05}%
  \BibitemOpen
  \bibfield  {author} {\bibinfo {author} {\bibfnamefont {A.}~\bibnamefont
  {Semenov}}, \bibinfo {author} {\bibfnamefont {A.}~\bibnamefont {Engel}},
  \bibinfo {author} {\bibfnamefont {H.-W.}\ \bibnamefont {Hubers}}, \bibinfo
  {author} {\bibfnamefont {K.}~\bibnamefont {Il'in}}, \ and\ \bibinfo {author}
  {\bibfnamefont {M.}~\bibnamefont {Siegel}},\ }\href@noop {} {\bibfield
  {journal} {\bibinfo  {journal} {Euro. Phys. J. B}\ }\textbf {\bibinfo
  {volume} {47}},\ \bibinfo {pages} {495} (\bibinfo {year} {2005})}\BibitemShut
  {NoStop}%
\bibitem [{\citenamefont {Gaudio}\ \emph {et~al.}(2014)\citenamefont {Gaudio},
  \citenamefont {op~'t Hoog}, \citenamefont {Zhou}, \citenamefont {Sahin},\
  and\ \citenamefont {Fiore}}]{Gaudio14}%
  \BibitemOpen
  \bibfield  {author} {\bibinfo {author} {\bibfnamefont {R.}~\bibnamefont
  {Gaudio}}, \bibinfo {author} {\bibfnamefont {K.}~\bibnamefont {op~'t Hoog}},
  \bibinfo {author} {\bibfnamefont {Z.}~\bibnamefont {Zhou}}, \bibinfo {author}
  {\bibfnamefont {D.}~\bibnamefont {Sahin}}, \ and\ \bibinfo {author}
  {\bibfnamefont {A.}~\bibnamefont {Fiore}},\ }\href@noop {} {\bibfield
  {journal} {\bibinfo  {journal} {Appl. Phys. Lett.}\ }\textbf {\bibinfo
  {volume} {105}},\ \bibinfo {pages} {222602} (\bibinfo {year}
  {2014})}\BibitemShut {NoStop}%
\bibitem [{\citenamefont {Charaev}\ \emph {et~al.}(2016)\citenamefont
  {Charaev}, \citenamefont {Semenov}, \citenamefont {Doerner}, \citenamefont
  {Gomard}, \citenamefont {Il'in},\ and\ \citenamefont {Siegel}}]{Charaev16}%
  \BibitemOpen
  \bibfield  {author} {\bibinfo {author} {\bibfnamefont {I.}~\bibnamefont
  {Charaev}}, \bibinfo {author} {\bibfnamefont {A.}~\bibnamefont {Semenov}},
  \bibinfo {author} {\bibfnamefont {S.}~\bibnamefont {Doerner}}, \bibinfo
  {author} {\bibfnamefont {G.}~\bibnamefont {Gomard}}, \bibinfo {author}
  {\bibfnamefont {K.}~\bibnamefont {Il'in}}, \ and\ \bibinfo {author}
  {\bibfnamefont {M.}~\bibnamefont {Siegel}},\ }\href@noop {} {\bibfield
  {journal} {\bibinfo  {journal} {Superconductor Science and Technology}\
  }\textbf {\bibinfo {volume} {30}},\ \bibinfo {pages} {025016} (\bibinfo
  {year} {2016})}\BibitemShut {NoStop}%
\bibitem [{\citenamefont {Renema}\ \emph {et~al.}(2014)\citenamefont {Renema},
  \citenamefont {Gaudio}, \citenamefont {Wang}, \citenamefont {Zhou},
  \citenamefont {Gaggero}, \citenamefont {Mattioli}, \citenamefont {Leoni},
  \citenamefont {Sahin}, \citenamefont {de~Dood}, \citenamefont {Fiore},\ and\
  \citenamefont {van Exter}}]{Renema14}%
  \BibitemOpen
  \bibfield  {author} {\bibinfo {author} {\bibfnamefont {J.~J.}\ \bibnamefont
  {Renema}}, \bibinfo {author} {\bibfnamefont {R.}~\bibnamefont {Gaudio}},
  \bibinfo {author} {\bibfnamefont {Q.}~\bibnamefont {Wang}}, \bibinfo {author}
  {\bibfnamefont {Z.}~\bibnamefont {Zhou}}, \bibinfo {author} {\bibfnamefont
  {A.}~\bibnamefont {Gaggero}}, \bibinfo {author} {\bibfnamefont
  {F.}~\bibnamefont {Mattioli}}, \bibinfo {author} {\bibfnamefont
  {R.}~\bibnamefont {Leoni}}, \bibinfo {author} {\bibfnamefont
  {D.}~\bibnamefont {Sahin}}, \bibinfo {author} {\bibfnamefont {M.~J.~A.}\
  \bibnamefont {de~Dood}}, \bibinfo {author} {\bibfnamefont {A.}~\bibnamefont
  {Fiore}}, \ and\ \bibinfo {author} {\bibfnamefont {M.~P.}\ \bibnamefont {van
  Exter}},\ }\href@noop {} {\bibfield  {journal} {\bibinfo  {journal} {Phys.
  Rev. Lett.}\ }\textbf {\bibinfo {volume} {112}},\ \bibinfo {pages} {117604}
  (\bibinfo {year} {2014})}\BibitemShut {NoStop}%
\bibitem [{\citenamefont {Clem}\ and\ \citenamefont {Berggren}(2011)}]{Clem11}%
  \BibitemOpen
  \bibfield  {author} {\bibinfo {author} {\bibfnamefont {J.~R.}\ \bibnamefont
  {Clem}}\ and\ \bibinfo {author} {\bibfnamefont {K.~K.}\ \bibnamefont
  {Berggren}},\ }\href@noop {} {\bibfield  {journal} {\bibinfo  {journal}
  {Phys. Rev. B}\ }\textbf {\bibinfo {volume} {84}},\ \bibinfo {pages} {174510}
  (\bibinfo {year} {2011})}\BibitemShut {NoStop}%
\bibitem [{\citenamefont {Hortensius}\ \emph {et~al.}(2012)\citenamefont
  {Hortensius}, \citenamefont {Driessen}, \citenamefont {Klapwijk},
  \citenamefont {Berggren},\ and\ \citenamefont {Clem}}]{Hortensius12}%
  \BibitemOpen
  \bibfield  {author} {\bibinfo {author} {\bibfnamefont {H.~L.}\ \bibnamefont
  {Hortensius}}, \bibinfo {author} {\bibfnamefont {E.~F.~C.}\ \bibnamefont
  {Driessen}}, \bibinfo {author} {\bibfnamefont {T.~M.}\ \bibnamefont
  {Klapwijk}}, \bibinfo {author} {\bibfnamefont {K.~K.}\ \bibnamefont
  {Berggren}}, \ and\ \bibinfo {author} {\bibfnamefont {J.~R.}\ \bibnamefont
  {Clem}},\ }\href@noop {} {\bibfield  {journal} {\bibinfo  {journal} {Appl.
  Phys. Lett.}\ }\textbf {\bibinfo {volume} {100}},\ \bibinfo {pages} {182602}
  (\bibinfo {year} {2012})}\BibitemShut {NoStop}%
\bibitem [{\citenamefont {Hortensius}\ \emph {et~al.}(2013)\citenamefont
  {Hortensius}, \citenamefont {Driessen},\ and\ \citenamefont
  {Klapwijk}}]{Hortensius13}%
  \BibitemOpen
  \bibfield  {author} {\bibinfo {author} {\bibfnamefont {H.~L.}\ \bibnamefont
  {Hortensius}}, \bibinfo {author} {\bibfnamefont {E.~F.~C.}\ \bibnamefont
  {Driessen}}, \ and\ \bibinfo {author} {\bibfnamefont {T.~M.}\ \bibnamefont
  {Klapwijk}},\ }\href@noop {} {\bibfield  {journal} {\bibinfo  {journal} {IEEE
  Trans. Appl. Supercond.}\ }\textbf {\bibinfo {volume} {23}},\ \bibinfo
  {pages} {2200705} (\bibinfo {year} {2013})}\BibitemShut {NoStop}%
\bibitem [{\citenamefont {Renema}\ \emph {et~al.}(2015)\citenamefont {Renema},
  \citenamefont {Wang}, \citenamefont {Gaudio}, \citenamefont {Komen},
  \citenamefont {op' Hoog}, \citenamefont {Sahin}, \citenamefont {Schilling},
  \citenamefont {van Exter}, \citenamefont {Fiore}, \citenamefont {Engel},\
  and\ \citenamefont {de~Dood}}]{Renema15}%
  \BibitemOpen
  \bibfield  {author} {\bibinfo {author} {\bibfnamefont {J.~J.}\ \bibnamefont
  {Renema}}, \bibinfo {author} {\bibfnamefont {Q.}~\bibnamefont {Wang}},
  \bibinfo {author} {\bibfnamefont {R.}~\bibnamefont {Gaudio}}, \bibinfo
  {author} {\bibfnamefont {I.}~\bibnamefont {Komen}}, \bibinfo {author}
  {\bibfnamefont {K.}~\bibnamefont {op' Hoog}}, \bibinfo {author}
  {\bibfnamefont {D.}~\bibnamefont {Sahin}}, \bibinfo {author} {\bibfnamefont
  {A.}~\bibnamefont {Schilling}}, \bibinfo {author} {\bibfnamefont {M.~P.}\
  \bibnamefont {van Exter}}, \bibinfo {author} {\bibfnamefont {A.}~\bibnamefont
  {Fiore}}, \bibinfo {author} {\bibfnamefont {A.}~\bibnamefont {Engel}}, \ and\
  \bibinfo {author} {\bibfnamefont {M.~J.~A.}\ \bibnamefont {de~Dood}},\
  }\href@noop {} {\bibfield  {journal} {\bibinfo  {journal} {Nano Lett.}\
  }\textbf {\bibinfo {volume} {15}},\ \bibinfo {pages} {4541} (\bibinfo {year}
  {2015})}\BibitemShut {NoStop}%
\bibitem [{\citenamefont {Vodolazov}\ \emph {et~al.}(2015)\citenamefont
  {Vodolazov}, \citenamefont {Korneeva}, \citenamefont {Semenov}, \citenamefont
  {Korneev},\ and\ \citenamefont {Goltsman}}]{Vodolazov15}%
  \BibitemOpen
  \bibfield  {author} {\bibinfo {author} {\bibfnamefont {D.~Y.}\ \bibnamefont
  {Vodolazov}}, \bibinfo {author} {\bibfnamefont {Y.~P.}\ \bibnamefont
  {Korneeva}}, \bibinfo {author} {\bibfnamefont {A.~V.}\ \bibnamefont
  {Semenov}}, \bibinfo {author} {\bibfnamefont {A.~A.}\ \bibnamefont
  {Korneev}}, \ and\ \bibinfo {author} {\bibfnamefont {G.~N.}\ \bibnamefont
  {Goltsman}},\ }\href@noop {} {\bibfield  {journal} {\bibinfo  {journal}
  {Phys. Rev. B}\ }\textbf {\bibinfo {volume} {92}},\ \bibinfo {pages} {104503}
  (\bibinfo {year} {2015})}\BibitemShut {NoStop}%
\bibitem [{\citenamefont {Engel}\ \emph
  {et~al.}(2015{\natexlab{a}})\citenamefont {Engel}, \citenamefont {Lonsky},
  \citenamefont {Zhang},\ and\ \citenamefont {Schilling}}]{Engel15}%
  \BibitemOpen
  \bibfield  {author} {\bibinfo {author} {\bibfnamefont {A.}~\bibnamefont
  {Engel}}, \bibinfo {author} {\bibfnamefont {J.}~\bibnamefont {Lonsky}},
  \bibinfo {author} {\bibfnamefont {X.}~\bibnamefont {Zhang}}, \ and\ \bibinfo
  {author} {\bibfnamefont {A.}~\bibnamefont {Schilling}},\ }\href@noop {}
  {\bibfield  {journal} {\bibinfo  {journal} {IEEE Trans. Appl. Supercond.}\
  }\textbf {\bibinfo {volume} {25}},\ \bibinfo {pages} {2200407} (\bibinfo
  {year} {2015}{\natexlab{a}})}\BibitemShut {NoStop}%
\bibitem [{\citenamefont {Engel}\ \emph
  {et~al.}(2015{\natexlab{b}})\citenamefont {Engel}, \citenamefont {Renema},
  \citenamefont {Il'in},\ and\ \citenamefont {Semenov}}]{Engel15review}%
  \BibitemOpen
  \bibfield  {author} {\bibinfo {author} {\bibfnamefont {A.}~\bibnamefont
  {Engel}}, \bibinfo {author} {\bibfnamefont {J.}~\bibnamefont {Renema}},
  \bibinfo {author} {\bibfnamefont {K.}~\bibnamefont {Il'in}}, \ and\ \bibinfo
  {author} {\bibfnamefont {A.}~\bibnamefont {Semenov}},\ }\href@noop {}
  {\bibfield  {journal} {\bibinfo  {journal} {Superconductor Science and
  Technology}\ }\textbf {\bibinfo {volume} {28}},\ \bibinfo {pages} {114003}
  (\bibinfo {year} {2015}{\natexlab{b}})}\BibitemShut {NoStop}%
\bibitem [{\citenamefont {Divochiy}\ \emph {et~al.}(2008)\citenamefont
  {Divochiy}, \citenamefont {Marsili}, \citenamefont {Bitauld}, \citenamefont
  {Gaggero}, \citenamefont {Leoni}, \citenamefont {Mattioli}, \citenamefont
  {Korneev}, \citenamefont {Seleznev}, \citenamefont {Kaurova}, \citenamefont
  {Minaeva}, \citenamefont {Gol'tsman}, \citenamefont {Lagoudakis},
  \citenamefont {Benkhaoul}, \citenamefont {Levy},\ and\ \citenamefont
  {Fiore}}]{Divochiy08}%
  \BibitemOpen
  \bibfield  {author} {\bibinfo {author} {\bibfnamefont {A.}~\bibnamefont
  {Divochiy}}, \bibinfo {author} {\bibfnamefont {F.}~\bibnamefont {Marsili}},
  \bibinfo {author} {\bibfnamefont {D.}~\bibnamefont {Bitauld}}, \bibinfo
  {author} {\bibfnamefont {A.}~\bibnamefont {Gaggero}}, \bibinfo {author}
  {\bibfnamefont {R.}~\bibnamefont {Leoni}}, \bibinfo {author} {\bibfnamefont
  {F.}~\bibnamefont {Mattioli}}, \bibinfo {author} {\bibfnamefont
  {A.}~\bibnamefont {Korneev}}, \bibinfo {author} {\bibfnamefont
  {V.}~\bibnamefont {Seleznev}}, \bibinfo {author} {\bibfnamefont
  {N.}~\bibnamefont {Kaurova}}, \bibinfo {author} {\bibfnamefont
  {O.}~\bibnamefont {Minaeva}}, \bibinfo {author} {\bibfnamefont
  {G.}~\bibnamefont {Gol'tsman}}, \bibinfo {author} {\bibfnamefont {K.~G.}\
  \bibnamefont {Lagoudakis}}, \bibinfo {author} {\bibfnamefont
  {M.}~\bibnamefont {Benkhaoul}}, \bibinfo {author} {\bibfnamefont
  {F.}~\bibnamefont {Levy}}, \ and\ \bibinfo {author} {\bibfnamefont
  {A.}~\bibnamefont {Fiore}},\ }\href@noop {} {\bibfield  {journal} {\bibinfo
  {journal} {Nat. Photonics}\ }\textbf {\bibinfo {volume} {2}},\ \bibinfo
  {pages} {302} (\bibinfo {year} {2008})}\BibitemShut {NoStop}%
\bibitem [{\citenamefont {Marsili}\ \emph
  {et~al.}(2011{\natexlab{b}})\citenamefont {Marsili}, \citenamefont {Najafi},
  \citenamefont {Dauler}, \citenamefont {Bellei}, \citenamefont {Hu},
  \citenamefont {Csete}, \citenamefont {Molnar},\ and\ \citenamefont
  {Berggren}}]{Marsili11PhtnRes}%
  \BibitemOpen
  \bibfield  {author} {\bibinfo {author} {\bibfnamefont {F.}~\bibnamefont
  {Marsili}}, \bibinfo {author} {\bibfnamefont {F.}~\bibnamefont {Najafi}},
  \bibinfo {author} {\bibfnamefont {E.}~\bibnamefont {Dauler}}, \bibinfo
  {author} {\bibfnamefont {F.}~\bibnamefont {Bellei}}, \bibinfo {author}
  {\bibfnamefont {X.}~\bibnamefont {Hu}}, \bibinfo {author} {\bibfnamefont
  {M.}~\bibnamefont {Csete}}, \bibinfo {author} {\bibfnamefont {R.~J.}\
  \bibnamefont {Molnar}}, \ and\ \bibinfo {author} {\bibfnamefont {K.~K.}\
  \bibnamefont {Berggren}},\ }\href@noop {} {\bibfield  {journal} {\bibinfo
  {journal} {Nano Lett.}\ }\textbf {\bibinfo {volume} {11}},\ \bibinfo {pages}
  {2048} (\bibinfo {year} {2011}{\natexlab{b}})}\BibitemShut {NoStop}%
\bibitem [{\citenamefont {Anant}\ \emph {et~al.}(2008)\citenamefont {Anant},
  \citenamefont {Kerman}, \citenamefont {Dauler}, \citenamefont {Yang},
  \citenamefont {Rosfjord},\ and\ \citenamefont {Berggren}}]{Anant08}%
  \BibitemOpen
  \bibfield  {author} {\bibinfo {author} {\bibfnamefont {V.}~\bibnamefont
  {Anant}}, \bibinfo {author} {\bibfnamefont {A.~J.}\ \bibnamefont {Kerman}},
  \bibinfo {author} {\bibfnamefont {E.~A.}\ \bibnamefont {Dauler}}, \bibinfo
  {author} {\bibfnamefont {J.~K.~W.}\ \bibnamefont {Yang}}, \bibinfo {author}
  {\bibfnamefont {K.~M.}\ \bibnamefont {Rosfjord}}, \ and\ \bibinfo {author}
  {\bibfnamefont {K.~K.}\ \bibnamefont {Berggren}},\ }\href@noop {} {\bibfield
  {journal} {\bibinfo  {journal} {Opt. Express}\ }\textbf {\bibinfo {volume}
  {16}},\ \bibinfo {pages} {10750} (\bibinfo {year} {2008})}\BibitemShut
  {NoStop}%
\bibitem [{\citenamefont {Driessen}\ \emph {et~al.}(2009)\citenamefont
  {Driessen}, \citenamefont {Braakman}, \citenamefont {Reiger}, \citenamefont
  {Dorenbos}, \citenamefont {Zwiller},\ and\ \citenamefont
  {de~Dood}}]{Driessen09Eur}%
  \BibitemOpen
  \bibfield  {author} {\bibinfo {author} {\bibfnamefont {E.~F.~C.}\
  \bibnamefont {Driessen}}, \bibinfo {author} {\bibfnamefont {F.}~\bibnamefont
  {Braakman}}, \bibinfo {author} {\bibfnamefont {E.}~\bibnamefont {Reiger}},
  \bibinfo {author} {\bibfnamefont {S.}~\bibnamefont {Dorenbos}}, \bibinfo
  {author} {\bibfnamefont {V.}~\bibnamefont {Zwiller}}, \ and\ \bibinfo
  {author} {\bibfnamefont {M.~J.~A.}\ \bibnamefont {de~Dood}},\ }\href@noop {}
  {\bibfield  {journal} {\bibinfo  {journal} {Eur. Phys. J. Appl. Phys.}\
  }\textbf {\bibinfo {volume} {47}},\ \bibinfo {pages} {10701} (\bibinfo {year}
  {2009})}\BibitemShut {NoStop}%
\bibitem [{\citenamefont {Zheng}\ \emph {et~al.}(2016)\citenamefont {Zheng},
  \citenamefont {Xu}, \citenamefont {Zhu}, \citenamefont {Jin}, \citenamefont
  {Kang}, \citenamefont {Xu}, \citenamefont {Chen},\ and\ \citenamefont
  {Wu}}]{Zheng16}%
  \BibitemOpen
  \bibfield  {author} {\bibinfo {author} {\bibfnamefont {F.}~\bibnamefont
  {Zheng}}, \bibinfo {author} {\bibfnamefont {R.}~\bibnamefont {Xu}}, \bibinfo
  {author} {\bibfnamefont {G.}~\bibnamefont {Zhu}}, \bibinfo {author}
  {\bibfnamefont {B.}~\bibnamefont {Jin}}, \bibinfo {author} {\bibfnamefont
  {L.}~\bibnamefont {Kang}}, \bibinfo {author} {\bibfnamefont {W.}~\bibnamefont
  {Xu}}, \bibinfo {author} {\bibfnamefont {J.}~\bibnamefont {Chen}}, \ and\
  \bibinfo {author} {\bibfnamefont {P.}~\bibnamefont {Wu}},\ }\href@noop {}
  {\bibfield  {journal} {\bibinfo  {journal} {Scientific Reports}\ }\textbf
  {\bibinfo {volume} {6}},\ \bibinfo {pages} {22710} (\bibinfo {year}
  {2016})}\BibitemShut {NoStop}%
\bibitem [{\citenamefont {RSoft}()}]{RSoft}%
  \BibitemOpen
  \bibfield  {author} {\bibinfo {author} {\bibnamefont {RSoft}},\ }\href@noop
  {} {}\bibinfo {howpublished}
  {\url{http://optics.synopsys.com/rsoft/}}\BibitemShut {NoStop}%
\bibitem [{\citenamefont {Jackson}(1983)}]{Jackson}%
  \BibitemOpen
  \bibfield  {author} {\bibinfo {author} {\bibfnamefont {J.~D.}\ \bibnamefont
  {Jackson}},\ }\href@noop {} {\emph {\bibinfo {title} {Classical
  Electrodynamics}}}\ (\bibinfo  {publisher} {Wiley $\&$ Sons},\ \bibinfo
  {address} {New York},\ \bibinfo {year} {1983})\BibitemShut {NoStop}%

\bibitem [{\citenamefont {Kozorezov}\ \emph {et~al.}(2015)\citenamefont
  {Kozorezov}, \citenamefont {Lambert}, \citenamefont {Marsili}, \citenamefont
  {Stevens}, \citenamefont {Verma}, \citenamefont {Stern}, \citenamefont
  {Horansky}, \citenamefont {Dyer}, \citenamefont {Duff}, \citenamefont
  {Pappas}, \citenamefont {Lita}, \citenamefont {Shaw}, \citenamefont {Mirin},\
  and\ \citenamefont {Nam}}]{Kozorezov15}%
  \BibitemOpen
  \bibfield  {author} {\bibinfo {author} {\bibfnamefont {A.~G.}\ \bibnamefont
  {Kozorezov}}, \bibinfo {author} {\bibfnamefont {C.}~\bibnamefont {Lambert}},
  \bibinfo {author} {\bibfnamefont {F.}~\bibnamefont {Marsili}}, \bibinfo
  {author} {\bibfnamefont {M.~J.}\ \bibnamefont {Stevens}}, \bibinfo {author}
  {\bibfnamefont {V.~B.}\ \bibnamefont {Verma}}, \bibinfo {author}
  {\bibfnamefont {J.~A.}\ \bibnamefont {Stern}}, \bibinfo {author}
  {\bibfnamefont {R.}~\bibnamefont {Horansky}}, \bibinfo {author}
  {\bibfnamefont {S.}~\bibnamefont {Dyer}}, \bibinfo {author} {\bibfnamefont
  {S.}~\bibnamefont {Duff}}, \bibinfo {author} {\bibfnamefont {D.~P.}\
  \bibnamefont {Pappas}}, \bibinfo {author} {\bibfnamefont {A.}~\bibnamefont
  {Lita}}, \bibinfo {author} {\bibfnamefont {M.~D.}\ \bibnamefont {Shaw}},
  \bibinfo {author} {\bibfnamefont {R.~P.}\ \bibnamefont {Mirin}}, \ and\
  \bibinfo {author} {\bibfnamefont {S.~W.}\ \bibnamefont {Nam}},\ }\href@noop
  {} {\bibfield  {journal} {\bibinfo  {journal} {Phys. Rev. B}\ }\textbf
  {\bibinfo {volume} {92}},\ \bibinfo {pages} {064504} (\bibinfo {year}
  {2015})}\BibitemShut {NoStop}%


\bibitem [{\citenamefont {Caloz}\ \emph {et~al.}(2017)\citenamefont {Caloz},
  \citenamefont {Korzh}, \citenamefont {Timoney}, \citenamefont {Weiss},
  \citenamefont {S.Gariglio}, \citenamefont {Warburton}, \citenamefont
  {Sch\"onenberger}, \citenamefont {Renema}, \citenamefont {Zbinden},\ and\
  \citenamefont {Bussieres}}]{Caloz16}%
  \BibitemOpen
  \bibfield  {author} {\bibinfo {author} {\bibfnamefont {M.}~\bibnamefont
  {Caloz}}, \bibinfo {author} {\bibfnamefont {B.}~\bibnamefont {Korzh}},
  \bibinfo {author} {\bibfnamefont {N.}~\bibnamefont {Timoney}}, \bibinfo
  {author} {\bibfnamefont {M.}~\bibnamefont {Weiss}}, \bibinfo {author}
  {\bibnamefont {S.Gariglio}}, \bibinfo {author} {\bibfnamefont
  {R.}~\bibnamefont {Warburton}}, \bibinfo {author} {\bibfnamefont
  {C.}~\bibnamefont {Sch\"onenberger}}, \bibinfo {author} {\bibfnamefont
  {J.}~\bibnamefont {Renema}}, \bibinfo {author} {\bibfnamefont
  {H.}~\bibnamefont {Zbinden}}, \ and\ \bibinfo {author} {\bibfnamefont
  {F.}~\bibnamefont {Bussieres}},\ }\href@noop {} {\bibfield  {journal}
  {\bibinfo  {journal} {Appl. Phys. Lett.}\ }\textbf {\bibinfo {volume}
  {110}},\ \bibinfo {pages} {083106} (\bibinfo {year} {2017})}\BibitemShut
  {NoStop}%


\bibitem [{\citenamefont {Renema}(2015)}]{RenemaThesis}%
  \BibitemOpen
  \bibfield  {author} {\bibinfo {author} {\bibfnamefont {J.~J.}\ \bibnamefont
  {Renema}},\ }\enquote {\bibinfo {title} {The physics of nanowire
  superconducting single-photon detectors},}\ \ (\bibinfo  {publisher} {Leiden
  University},\ \bibinfo {address} {Leiden, The Netherlands},\ \bibinfo {year}
  {2015})\ p.~\bibinfo {pages} {81}\BibitemShut {NoStop}%

\bibitem [{\citenamefont {Dodge}\ \emph {et~al.}(1969)\citenamefont {Dodge},
  \citenamefont {Malitson},\ and\ \citenamefont {Mahan}}]{Dodge69}%
  \BibitemOpen
  \bibfield  {author} {\bibinfo {author} {\bibfnamefont {M.~J.}\ \bibnamefont
  {Dodge}}, \bibinfo {author} {\bibfnamefont {I.~H.}\ \bibnamefont {Malitson}},
  \ and\ \bibinfo {author} {\bibfnamefont {A.~I.}\ \bibnamefont {Mahan}},\
  }\href@noop {} {\bibfield  {journal} {\bibinfo  {journal} {Appl. Opt.}\
  }\textbf {\bibinfo {volume} {8}},\ \bibinfo {pages} {1703} (\bibinfo {year}
  {1969})}\BibitemShut {NoStop}%
\bibitem [{\citenamefont {Sahin}(2014)}]{SahinThesis}%
  \BibitemOpen
  \bibfield  {author} {\bibinfo {author} {\bibfnamefont {D.}~\bibnamefont
  {Sahin}},\ }\emph {\bibinfo {title} {Waveguide single-photon and
  photon-number resolving detectors}},\ \href@noop {} {Ph.D. thesis},\ \bibinfo
   {school} {Eindhoven University of Technology}, \bibinfo {address}
  {Eindhoven, The Netherlands} (\bibinfo {year} {2014})\BibitemShut {NoStop}%
\bibitem [{\citenamefont {Heath}\ \emph {et~al.}(2015)\citenamefont {Heath},
  \citenamefont {Tanner}, \citenamefont {Drysdale}, \citenamefont {Miki},
  \citenamefont {Giannini}, \citenamefont {Maier},\ and\ \citenamefont
  {Hadfield†}}]{Heath15}%
  \BibitemOpen
  \bibfield  {author} {\bibinfo {author} {\bibfnamefont {R.}~\bibnamefont
  {Heath}}, \bibinfo {author} {\bibfnamefont {M.}~\bibnamefont {Tanner}},
  \bibinfo {author} {\bibfnamefont {T.}~\bibnamefont {Drysdale}}, \bibinfo
  {author} {\bibfnamefont {S.}~\bibnamefont {Miki}}, \bibinfo {author}
  {\bibfnamefont {V.}~\bibnamefont {Giannini}}, \bibinfo {author}
  {\bibfnamefont {S.}~\bibnamefont {Maier}}, \ and\ \bibinfo {author}
  {\bibfnamefont {R.}~\bibnamefont {Hadfield†}},\ }\href@noop {} {\bibfield
  {journal} {\bibinfo  {journal} {Nano Lett.}\ }\textbf {\bibinfo {volume}
  {15}},\ \bibinfo {pages} {819} (\bibinfo {year} {2015})}\BibitemShut
  {NoStop}%
\bibitem [{\citenamefont {Green}\ and\ \citenamefont
  {Keevers}(1995)}]{Green95}%
  \BibitemOpen
  \bibfield  {author} {\bibinfo {author} {\bibfnamefont {M.~A.}\ \bibnamefont
  {Green}}\ and\ \bibinfo {author} {\bibfnamefont {M.~J.}\ \bibnamefont
  {Keevers}},\ }\href@noop {} {\bibfield  {journal} {\bibinfo  {journal}
  {Progress in Photovoltaics}\ }\textbf {\bibinfo {volume} {3}},\ \bibinfo
  {pages} {189} (\bibinfo {year} {1995})}\BibitemShut {NoStop}%

\bibitem [{\citenamefont {Hass}\ and\ \citenamefont {Salzberg}(1954)}]{Hass54}%
  \BibitemOpen
  \bibfield  {author} {\bibinfo {author} {\bibfnamefont {G.}~\bibnamefont
  {Hass}}\ and\ \bibinfo {author} {\bibfnamefont {C.~D.}\ \bibnamefont
  {Salzberg}},\ }\href@noop {} {\bibfield  {journal} {\bibinfo  {journal} {J.
  Opt. Soc. Am.}\ }\textbf {\bibinfo {volume} {44}},\ \bibinfo {pages} {181}
  (\bibinfo {year} {1954})}\BibitemShut {NoStop}%

\bibitem [{\citenamefont {Renema}\ \emph {et~al.}(2013)\citenamefont {Renema},
  \citenamefont {Frucci}, \citenamefont {Zhou}, \citenamefont {Mattioli},
  \citenamefont {Gaggero}, \citenamefont {Leoni}, \citenamefont {de~Dood},
  \citenamefont {Fiore},\ and\ \citenamefont {van Exter}}]{Renema13}%
  \BibitemOpen
  \bibfield  {author} {\bibinfo {author} {\bibfnamefont {J.~J.}\ \bibnamefont
  {Renema}}, \bibinfo {author} {\bibfnamefont {G.}~\bibnamefont {Frucci}},
  \bibinfo {author} {\bibfnamefont {Z.}~\bibnamefont {Zhou}}, \bibinfo {author}
  {\bibfnamefont {F.}~\bibnamefont {Mattioli}}, \bibinfo {author}
  {\bibfnamefont {A.}~\bibnamefont {Gaggero}}, \bibinfo {author} {\bibfnamefont
  {R.}~\bibnamefont {Leoni}}, \bibinfo {author} {\bibfnamefont {M.~J.~A.}\
  \bibnamefont {de~Dood}}, \bibinfo {author} {\bibfnamefont {A.}~\bibnamefont
  {Fiore}}, \ and\ \bibinfo {author} {\bibfnamefont {M.~P.}\ \bibnamefont {van
  Exter}},\ }\href@noop {} {\bibfield  {journal} {\bibinfo  {journal} {Phys.
  Rev. B}\ }\textbf {\bibinfo {volume} {87}},\ \bibinfo {pages} {174526}
  (\bibinfo {year} {2013})}\BibitemShut {NoStop}%
\bibitem [{\citenamefont {Wang}\ \emph {et~al.}(2015)\citenamefont {Wang},
  \citenamefont {Renema}, \citenamefont {Engel}, \citenamefont {van Exter},\
  and\ \citenamefont {de~Dood}}]{Wang15}%
  \BibitemOpen
  \bibfield  {author} {\bibinfo {author} {\bibfnamefont {Q.}~\bibnamefont
  {Wang}}, \bibinfo {author} {\bibfnamefont {J.}~\bibnamefont {Renema}},
  \bibinfo {author} {\bibfnamefont {A.}~\bibnamefont {Engel}}, \bibinfo
  {author} {\bibfnamefont {M.}~\bibnamefont {van Exter}}, \ and\ \bibinfo
  {author} {\bibfnamefont {M.}~\bibnamefont {de~Dood}},\ }\href@noop {}
  {\bibfield  {journal} {\bibinfo  {journal} {Opt. Expr.}\ }\textbf {\bibinfo
  {volume} {23}},\ \bibinfo {pages} {24873} (\bibinfo {year}
  {2015})}\BibitemShut {NoStop}%
\end{thebibliography}
\end{document}